\documentclass[aps,pre,preprint,groupedaddress,showpacs]{revtex4-1}
\usepackage{graphicx,amsmath}
\pdfoutput=1
\begin{document}

\title{Notes about the Macroscopic Fluctuating Theory}


\author{P.L. Garrido}
\email[]{garrido@onsager.ugr.es}
\affiliation{Instituto Carlos I de F{\'\i}sica Te{\'o}rica y Computacional. Universidad de Granada. E-18071 Granada. Spain }

\date{\today}
\begin{abstract}
The Macroscopic Fluctuating Theory developed during the last thirty years is applied to generic systems described by continuum fields $\phi(x,t)$ that evolve by a Langevin equation that locally either conserves or not the field. The aim of this paper is to present well known concepts and results following a general framework in a practical and self consistent way.
From the probability of a path we study general properties of the system's stationary state. In particular we focus on the study of the quasi-potential that defines the stationary distribution at the small noise limit. It is derived the system's {\it adjoint dynamics} that it is the time reversal Markov process of the system. The equilibrium is assumed to be the unique stationary state that is dynamically time reversal and therefore it can be reached by the system only when the adjoint dynamics is equal to the original one. This condition characterizes the dynamics of systems with  equilibrium from the ones with nonequilibrium stationary states. 
That property is confronted with the {\it macroscopic reversibility} that occurs when the most probable path to create a fluctuation from the stationary state is equal to the time reversed path that relaxes it. The lack of this symmetry implies a nonequilibrium stationary state however the converse isn't true.  Finally we study extensively the two-body correlations at the stationary state and we derive some generic properties at a variety of situations, including a discussion about the equivalence of ensembles in non-equilibrium systems. 
\end{abstract}
\maketitle

\section{Introduction}
 Nonequilibrium stationary states appear when external agents or boundary conditions drive, an otherwise system at equilibrium, out of it by pumping and extracting continuously  particles, energy.....  or simply when a microscopic dynamics breaks some spatial symmetry and imposes long range correlations between their components. In fact one should recognize that nonequilibrium stationary states are ubiquitous in Nature and in contrast, equilibrium states very scarce and difficult to observe. 
 It is curious that for equilibrium states we have the Thermodynamics and the Boltzmann-Gibbs ensemble theories that permits us to understand and predict the system's macroscopic and mesoscopic (fluctuating) behavior with a great success. 
 However there are no similar general theories  for systems at nonequilibrium stationary states. Therefore it is a hard task  to get some general result, to derive any prediction to be checked by experiments (numerical or not) or just to reproduce some observation.
After more than two centuries of research we have just a few (but very relevant from a practical point of view)  phenomenological macroscopic  relations as for instance the Navier-Stokes Equation or the Fourier's Law that are used to describe the behavior of a  given fluid or the heat transport respectively \cite{Bat}. These depend on constitutive parameters that are obtained by experiments. Moreover they are based in the assumption that some equilibrium thermodynamic relations apply locally in the system assuming in some sense that the system is not far from equilibrium \cite{Groot}. Finally, these nonequilibrium descriptions do not include any mesoscopic behavior and when needed it is added by hand by taking again as reference the fluctuations at equilibrium \cite{Sengers}. 
There has been been a lot of effort in the last decades to connect microscopic models with these phenomenological equations \cite{Spohn}. This strategy is very important in order to unveil the properties of the free parameters (viscosity, thermal conductivity,..) and the range of applicability of such relations. Nevertheless it is only partially resolved the rigorous derivation of hydrodynamic (macroscopic) equations starting  from a Boltzmann equation as the microscopic description \cite{Ros}. 

From a more fundamental side in the last forty years there have been several relevant achievements for systems at nonequilibrium stationary states in the  understanding of  the generic structure of their stationary attractor in the phase space and into the definition of the invariant measures on it  \cite{Gallavotti3}. This Boltzmann-like strategy has permitted for instance to introduce the SRB measures as typical for these systems \cite{Ruelle} or to derive the Fluctuation Theorem \cite{Cohen}. However from a practical level we are still far from a truly useful nonequilibrium ensemble theory.

Despite all the commented efforts, maybe there are some more advances for a complete theory of nonequilibrium stationary states when we get a deeply understanding of concrete models and systems.
There are many techniques, theoretical approaches and/or computer simulated models that permits to get some insight of some particular nonequilibrium model in physics, ecology, biophysics,... Each of the studies present different characterizations of their nonequilibrium stationary state and it is measured a variety of observables that are assumed to be the optimal ones to describe the particular phenomena. In our opinion, the main problem nowadays is to find a common theoretical framework flexible enough that could permit us to apply it to different situations. That would permit us to compare different approaches, results or ideas in trying to find the essential properties that characterize nonequilibrium stationary states.

A path in that direction was done by R. Graham and co-workers by studying the stationary distribution of a generic white noise stochastic Langevin equation with finite degrees of freedom in the small noise limit \cite{seminal}. They find the integrability conditions for the existence of a quasi-potential that is the nonequilibrium version of the free energy functionals in thermodynamics. Moreover, they analyze some explicit examples with two degrees of freedom \cite{Graham_Tel0} or how to deal with stationary states with several attractors \cite{Graham_Tel1}. The reader will find in our paper a translation of some of their results and we refer to their seminal work for a deeper understanding of some concepts that, in the spirit of this paper, we just touch lightly.

In the same line of though  a very interesting effort has been done by Bertini at al. \cite{Bertini0} during the last years  by developping a mesoscopic theory for diffusive systems that they call {\it Macroscopic Fluctuating Theory} (MFT). MFT is based in three main assumptions. First, the existence of a well defined  hydrodynamic (macroscopic) description of the system.  Second,  it is assumed that the fluctuating behavior of the macroscopic variables follows a Large Deviation Principle. And third,  it is used  the {\it  Fundamental Principle} (as they call it) that it is a kind of generalized {\it detailed balance condition} that connects the way the system relax from a fluctuation to how that was created. All those assumptions are based in previous rigorous results in some one dimensional microscopic stochastic nonequilibrium models as for example the boundary driven Symmetric Simple Exclusion Process, the Weakly Asymetric Exclusion Process or the Kipnis Marchioro Presutti model (see for instance the review by Bertini et al. \cite{Bertini}).  From this solid starting point MFT intends to obtain general properties of the stationary state on a variety of systems in a very serious attempt to globally understand the behavior of diffusive systems  from a theoretical point of view. In practice, one may say that MFT is the application of Graham work to diffusive systems with infinity degrees of freedom or, in other words, systems described by Langevin equations for fields.

In this paper we extend (in a non-rigorous way) all these seminal work to more general nonequilibrium systems. In order to do that, we assume as starting point that our system is defined at the mesoscopic level by a  Langevin equation for fields with a local white noise field that it is uncorrelated in time.  This set up allows us  to apply many of the MFT concepts general systems. In fact we have studied dynamics that locally do not conserve the fields and dynamics that conserve them. The later include the diffusive ones where we reproduce again the known results from MFT.  
In this paper we focus in the one component continuum Langevin equations with conserved or non-conserved dynamics. In Section II we define the starting  equations, some notations and a set of basic definitions and relations. Section III is devoted to study the stationary probability distribution that in the small noise limit  is given by a functional of the fields that is called {\it quasi-potential}. We use the Path Probability  to obtain a Hamilton-Jacobi partial differential equation for the quasi-potential.  It can be formally solved by using the method of characteristics that give us the effective dynamic equations describing the most probable path to create a given fluctuation. That permits us to find some general properties for the quasi-potential.   We introduce in Section IV the macroscopic reversibility property that it is defined when paths to create a fluctuation  and the time reversed deterministic path to relax it coincide. We argue that a system  that is macroscopic reversible has a quasi-potential with existing and continuous first and second functional field derivatives. Nevertheless this property it is not enough to guarantee that the system stationary state is an equilibrium state. In Section V we introduce the Fundamental Principle that defines the system's {\it adjoint dynamics}  when the time is reversed  from a generalized detailed balance condition on paths. Typically the adjoint dynamics is different from the original one reflecting the fact that there exists some disipation that  can be related to the existence of a non-zero entropy production (see for instance \cite{LB}). In this context the equilibrium is the stationary state of the system when both dynamics coincide and that is consistent with the Onsager's idea that the microscopic reversibility extens to the mesoscopic level for systems at equilibrium. In Section V we study the spatial correlations at the stationary state. We obtain the general set of closed equations to study them for the conserved and non-conserved cases and we apply them to some well known situations in order to make explicit the power of this theoretical scheme. Moreover we build the conditions in which conserved and non-conserved situations develop the same quasi-potential around the stationary state. That is an attempt to build nonequilibrium dynamical ensembles. 

We are convinced that this MFT generalization, although do not introduce new paradigms in our basic knowledge of this problem, establish a theoretical framework that could be useful to many researchers in different fields were these continuous Langevin models are used. 

\section{Langevin description of mesoscopic systems}

We assume that our system, at a hydrodynamic level of description, is completely defined by a unique scalar field $\phi_D(x,t)\in {\rm I\!R}$ where $x\in \Lambda \subset {\rm I\!R}^d$, $d$ is the spatial dimension and $t$ is the time. In this paper we restrict ourselves to this case for the sake of simplicity but one can straightforward generalize all the results below to systems described by vector fields. The field evolution is solution of a nonlinear partial differential equation. Along this work we are going to consider two separate family of dynamics: the reaction dynamics (RD) that doesn't conserve the field locally, and the diffusive dynamics (DD) where the field is locally conserved under the evolution:
\begin{eqnarray}
\partial_t\phi_D(x,t)&=&F[\phi_D;x,t] \quad \text{(RD case)}\nonumber\\
 \partial\phi_D(x,t)&+&\nabla\cdot G[\phi_D;x,t]=0 \quad \text{(DD case)}\label{det}
\end{eqnarray}
where $F$ and  $G$ are given local  functionals of $\phi(x,t)$, $\nabla\phi(x,t)$, $\ldots$.
Our set of equations are solved typically for  given boundary conditions  $(\phi_D(x,t)=f(x), x\in\partial\Lambda, \forall t)$, and an initial state $(\phi_D(x,0)=\phi_0(x), x\in\Lambda)$. That determines (hopefully) the solution $\phi_D(x,t)$ that is also called  {\it Deterministic or Classical solution}. The {\it stationary state}, $\phi^*(x)=\lim_{t\rightarrow\infty}\phi_D(x,t)$, is  the stationary solution of the hydrodynamic equation:
\begin{equation}
F[\phi^*;x]=0\quad \text{(RD case)      or}  \quad \nabla\cdot G[\phi^*;x]=0 \quad \text{(DD case)}\label{se}
\end{equation}
We  assume for simplicity that $\phi^*$ is unique (in all cases). Observe that for RD and DD with open boundary conditions the stationary state is characterized by the model parameters and/or the boundary conditions. In a DD with periodic boundary conditons, the initial configuration fix the average density field on the system and  that it  is conserved by the dynamics and, therefore is a parameter that too determines the stationary state. It is also considered along this work that the dynamics are dynamically stable in the sense that all the eigenvalues for the linearized dynamics have a nonzero and negative real part. More precisely, let us expand the hydrodynamic equation around the stationary state: $\phi(x,t)=\phi^*(x)+\epsilon(x,t)$. Then we get
\begin{equation}
\partial_t\epsilon(x,t)=\int_{\Lambda}dy\,A(x,y)\epsilon(y) +O(\epsilon^2)
\end{equation}
where
\begin{equation}
A(x,y)=\frac{\delta F[\phi;x]}{\delta\phi(y)}\biggr\vert_{\phi=\phi^*}\quad \text{(RD case)  or}  \quad A(x,y)=\frac{\delta \nabla G[\phi;x]}{\delta\phi(y)}\biggr\vert_{\phi=\phi^*}\quad \text{(DD case)}\label{sta}
\end{equation}
then we assume that all the eigenvalues $\lambda$ of the operator $A$ which are solution of the equation $\det(A-{\rm I\!I}\lambda)=0$ are such that $Re(\lambda)<0$. This property guarantees that the stationary state is not time dependent. These are the class of stationary states we are going to study in this paper. Obviously there are many other stationary states but they are out the scope of this paper. 

The mesoscopic description is built from the hydrodynamics of the system by assuming that the system dynamics is given by a Langevin equation with a white noise: 
\begin{itemize}
\item {\it Reaction dynamics (RD):} The mesoscopic state of the system is described by the local field $\phi(x,t)$ that evolves by:
\begin{equation}
\partial_t\phi(x,t)=F[\phi;x,t]+h[\phi;x,t]\xi(x,t)\label{lan1}
\end{equation}
\item {\it Diffusion dynamics (DD):} The state of the system is completely described by the local field $\phi(x,t)$ and the local current $j(x,t)$. Their evolution is given by the continuity equation:
\begin{equation}
\partial_t\phi(x,t)+\nabla\cdot j(x,t)=0
\end{equation}
where the current $j$ is chosen to be of the form:
\begin{equation}
j_{\alpha}(x,t)=G_{\alpha}[\phi;x,t]+\sum_{\beta=1}^d\sigma_{\alpha,\beta}[\phi;x,t]\psi_{\beta}(x,t)  \quad \alpha=1,\ldots,d
\end{equation}
\end{itemize}
Here $G$ and $\sigma$ are given local  funcionals of $\phi(x,t)$. 

In both cases the boundary conditions and the initial state are the ones given for $\phi_D$. $\xi(x,t)$ and $\psi_{\alpha}(x,t)$ are uncorrelated gaussian random variables:
\begin{eqnarray}
\langle\xi(x,t)\rangle&=&0\nonumber\\
\langle\xi(x,t)\xi(x',t')\rangle&=&\frac{1}{\Omega}\delta(x-x')\delta(t-t')\nonumber\\
\langle\psi_{\alpha}(x,t)\rangle&=&0\nonumber\\
\langle\psi_{\alpha}(x,t)\psi_{\beta}(x',t')\rangle&=&\frac{1}{\Omega}\delta_{\alpha,\beta}\delta(x-x')\delta(t-t')
\end{eqnarray}
Observe that for the sake of simplicity the correlation matrix for the $\psi_\alpha$ noise is chosen to be proportional to the identity.
$\Omega>0$ is the parameter that controls the time and spatial separation between the mesoscopic and hydrodynamic descriptions. It is assumed that $\Omega$ is large and therefore, the fluctuations are  just perturbations to the {\it deterministic}  or macroscopic case ($\Omega\rightarrow\infty$). That is why we call  {\it Macroscopic Fluctuating Theory (MFT)}. As we will see, the assumption of small fluctuations does not imply small effects for a nonequilibrium stationary state. In fact the structure of many observables and potentials change  dramatically with respect to the equilibrium case even if we are very near to it. Therefore, even in the most simple cases, we should keep in mind that a nonequilibrum stationary state is going to be related with the existence of long range correlations and non-local and/or singular probability distributions. 

The presence of the random variables $\xi$ or $\psi_\alpha$ implies that the system evolution is characterized by probabilities. At this point it could be interesting to the reader to consult some classical books about stochastic phenomena for systems mainly with finite number of degrees of freedom. There it  can be find the general basic properties of these Langevin equations and  many applications to  simple illustrative cases \cite{Gardiner}. In particular we are interested in the probability that the system follows a given evolution path $\phi$ in a time interval. From the Langevin equations we can explicitly construct such probability that contains most of the interesting physics of the system. 

 An arbitrary path is defined by the set:  $\left\{\phi\right\}[t_0,t_1]=(\phi(x,t),x\in\Lambda, t\in[t_0,t_1])$. The probability of such path is just the sum over all set of random variables that recreate the path multiplied by their probability. For the RD case the path probability is given by:
\begin{eqnarray}
P[\left\{\phi\right\}[t_0,t_1]]&=&cte\int D\xi \exp\left[-\frac{\Omega}{2}\int_{-\infty}^{\infty}dt\int_{\Lambda}dx\,\xi(x,t)^2 \right]\cdot\nonumber\\
&&\prod_{t\in[t_0,t_1]}\prod_{x\in\Lambda}\delta\left(\partial_t\phi(x,t)-F[\phi;x,t]-h[\phi;x,t]\xi(x,t)\right)\label{path}
\end{eqnarray}
where the constant is found by the normalization of $P$. We use the integral representation of Dirac's delta to write
\begin{eqnarray}
P[\left\{\phi\right\}[t_0,t_1]]&=&cte\int D\pi\exp\left[-\Omega\int_{t_0}^{t_1}dt\int_\Lambda dx\,\pi(x,t)\left(\partial_t\phi(x,t)-F[\phi;x,t]\right)\right]\cdot\nonumber\\
&&\int D\xi\exp\left[[-\Omega\int_{t_0}^{t_1}dt\int_{\Lambda}dx\,\left(\frac{1}{2}\xi(x,t)^2-\pi(x,t)h[\phi;x,t]\xi(x,t)\right)\right]
\end{eqnarray}
In this way we can do explicitly the gaussian integral over the noise variables and we get
\begin{equation}
P[\left\{\phi\right\}[t_0,t_1]]=cte\int D\pi \exp\left[-\Omega\int_{t_0}^{t_1}dt\left[\int_\Lambda dx\, \pi(x,t)\partial_t\phi(x,t)-H[\phi(t),\pi(t)]\right]\right] \label{pathg}
\end{equation}
where
\begin{equation}
H[\phi,\pi]=\int_\Lambda dx\,\pi(x)\left[F[\phi;x]+\frac{1}{2}\pi(x)h[\phi;x]^2\right]\label{hamRD}
\end{equation}
We could do explicitly the integral over $\pi$ but let us keep this expression for a later use.

For the DD case first we have to give the path probability for the variables that define the state of the system $(\phi,j)$:
\begin{eqnarray}
&&P[\left\{\phi,j\right\}[t_0,t_1]]=cte\int D\psi \exp\left[-\frac{\Omega}{2}\sum_\alpha\int_{-\infty}^{\infty}dt\int_{\Lambda}dx\,\psi_\alpha(x,t)^2 \right]\cdot\nonumber\\
&&\prod_{t\in[t_0,t_1]}\prod_{x\in\Lambda}\left[\delta\left(\partial_t\phi(x,t)+\nabla\cdot j(x,t)\right)\prod_\alpha\delta(j_\alpha(x,t)-G_\alpha[\phi;x,t]-\sum_{\beta}\sigma_{\alpha\beta}[\phi;x,t]\psi_\beta(x,t))\right]\nonumber\\
&&=cte \exp\left[-\frac{\Omega}{2}\int_{t_0}^{t_1}dt\int_{\Lambda}dx\,(j(x,t)-G[\phi;x,t])\cdot \chi^{-1}[\phi;x,t](j(x,t)-G[\phi;x,t])\right]\cdot\nonumber\\
&&\prod_{t\in[t_0,t_1]}\prod_{x\in\Lambda}\delta\left(\partial_t\phi(x,t)+\nabla\cdot j(x,t)\right)\label{path2}
\end{eqnarray}
where $\chi[\phi;x,t]$ is a symmetric invertible matrix with components 
\begin{equation}
\chi_{\alpha,\beta}[\phi;x,t]=\sum_{\gamma=1}^d\sigma_{\alpha,\gamma}[\phi;x,t]\sigma_{\beta,\gamma}[\phi;x,t]
\end{equation}
From this expression we can deduce the path probability for $\phi$:
\begin{equation}
P[\left\{\phi\right\}[t_0,t_1]]=\int Dj \,P[\left\{\phi,j\right\}[t_0,t_1]]
\end{equation}
Again, we  introduce the Dirac's delta representation, we take $\pi(x,t)=0\,\forall x\in\partial\Lambda$ and we integrate over the $j$'s.  After that we get the same expression eq.(\ref{pathg}) with
\begin{equation}
H[\phi,\pi]=\int_\Lambda dx\,\nabla\pi(x)\cdot\left[G[\phi;x]+\frac{1}{2}\chi[\phi;x]\nabla\pi(x)\right]\label{hamDD}
\end{equation}
We are ready to study the stationary distribution at the small noise limit ($\Omega\rightarrow\infty$).

\section{The Stationary State and the Quasipotential}

We assume in that from almost any initial condition the system evolves towards an unique stationary probability distribution $P_{st}[\phi]$. This stationary distribution could be computed by using the path probability \ref{pathg} or as the solution of the Fokker-Planck equation (see Appendix I) such that  $\partial_t P_{st}[\phi]=0$.  Let us focus in the first strategy because, as we will see, it includes a least action variational principle that in a more generic context is very important in order to select the absolute minimal solution when there are multiple local minimizers of the action (multiple solutions of the stationary Fokker-Planck equation).

The main idea to define the stationary distribution from the path probability  is to use the fact that the probability to go from a given starting state to another final in a time interval is just the sum of all the probabilities of every path that connects both states. Therefore the stationary probability to be at state $\eta$ is the probability to be at the stationary state $\phi^*$ times the probability to go from $\phi^*$ to $\eta$ in a infinite time interval:  
\begin{equation}
P_{st}[\eta]=P_{st}[\phi^*]\int D\phi\, P[\{\phi\}[-\infty,0]]\prod_{x\in\Lambda}\left[\delta(\phi(x,0)-\eta(x))\delta(\phi(x,-\infty)-\phi(x)^*)\right]\label{pspath}
\end{equation}
where $P[\{\phi\}[t_0,t_1]]$ is given by eq.(\ref{pathg}) and $H$ are obtained from  eqs. (\ref{hamRD}) or (\ref{hamDD}) for the RD or DD cases respectively. Except to some very simple models we do not know how to get explicit expressions for $P_{st}[\phi]$. However we can use the fact that the noise intensity is very small to significantly simplify the problem. When $\Omega\rightarrow\infty$  the stationary probability distribution is of the form:
\begin{equation}
P_{st}[\eta]\simeq \exp\left[-\Omega V_0[\eta]\right]\label{v0}
\end{equation}
where $V_0[\eta]$ is the so called {\it quasipotential}. Observe that in the strict limit $\Omega\rightarrow\infty$ we should get the stationary deterministic solution (\ref{se}). In other words:
\begin{equation}
P_{st}[\eta]=\prod_{x\in\Lambda}\delta(\eta(x)-\phi^*(x))
\end{equation}
This implies that $\phi^*(x)$ is the absolute minimum of the quasi potential: 
\begin{equation}
\frac{\delta V_0[\phi^*]}{\delta\phi^*(x)}=0\quad\forall x\in\Lambda
\end{equation}
When $\Omega\rightarrow\infty$ the integrals in eq.(\ref{pspath}) are dominated by the path $(\phi,\pi)$ that minimizes the argument of the exponential. This implies:
\begin{equation}
V_0[\eta]=V_0[\phi^*]+\inf_{\phi,\pi}\left\{
\int_{-\infty}^0 dt\left[\int_\Lambda dx \pi(x,t)\partial_t\phi(x,t)-H[\phi(t),\pi(t)]\right]
\right\}\label{action}
\end{equation}
with $H$ given by eqs. (\ref{hamRD}) and (\ref{hamDD}) for the RD and DD systems respectively. Moreover $\phi(x,-\infty)=\phi^*(x)$, $\phi(x,0)=\eta(x)$. $\pi(x,-\infty)=0$ and $H[\phi^*,0]=0$ by construction. This equation is just the Hamilton's variational principle with $H$ being the hamiltonian \cite{Goldstein}. Therefore, the fields $(\bar\pi(x,t),\bar\phi(x,t))$ that minimize the right hand side of eq.(\ref{action}) are solution of the Hamilton equations:
\begin{eqnarray}
\partial_t\bar\phi(x,t)&=&\frac{\delta H[\bar\phi(t),\bar\pi(t)]}{\delta\bar\pi(x,t)}\nonumber\\
\partial_t\bar\pi(x,t)&=&-\frac{\delta H[\bar\phi(t),\bar\pi(t)]}{\delta\bar\phi(x,t)}\label{Heq}
\end{eqnarray}
with the above boundary conditions. 
Observe that $H[\bar\phi(t),\bar\pi(t)]=0$ because the hamiltonian is constant along any trajectory and it is zero at the initial condition. Therefore eq.(\ref{action}) can be rewritten
\begin{equation}
V_0[\eta]=V_0[\phi^*]+
\int_{-\infty}^0 dt\int_\Lambda dx \bar\pi(x,t)\partial_t\bar\phi(x,t)\label{action2}
\end{equation}

At this point we stress some general issues:
\begin{itemize}
\item We can use the path probability to compute  the probability to observe a given value of the space and time averages of fields or functions of them in a time interval. For  enough large times we can use the Large Deviation Principle to get the probability to observe such averaged value and also some relations that are a kind of generalized Green-Kubo formulas for systems at nonequilibrium stationary states. This is a very intense research domain that intends to get general properties of these systems through the study of such distributions by theoretical approximations and computer simulations (see Appendix II).

\item We could also use the stationary Fokker-Planck equation, $\partial_t P_{st}[\phi]=0$ (see Appendix I) to do a $\Omega^{-1}$ expansion  on it assuming eq.(\ref{v0}). At the lowest order of such expansion we get the Hamilton-Jacobi equation:
\begin{equation}
H[\bar\phi,\frac{\delta V_0[\bar\phi]}{\delta\bar\phi}]=0\label{HJ}
\end{equation}
with $H$ given by eqs. (\ref{hamRD}) or (\ref{hamDD}) for the RD and DD systems respectively.
The equation can be solved by the method of characteristics which   implies at the end to solve the same Hamilton equations (\ref{Heq})  (see Appendix III). That is, $\bar\pi(x,t)=\delta V_0[\bar\phi]/\delta\bar\phi(x,t)$ $\forall x,t$ and, in particular,  $\delta V_0[\phi^*]/\delta\phi^*(x)=0$ when $t\rightarrow -\infty$ as expected. 

\item We can study the eq. (\ref{Heq}) near the initial condition $(\phi^*,0)$. The linear approximation is
\begin{eqnarray}
\partial_t\epsilon(x,t)&=&\int_\Lambda dy\left[ A(x,y)\epsilon(y,t)+B(x,y)\pi(x,t)\right]\nonumber\\
\partial_t\pi(x,t)&=&-\int_\Lambda dy A(y,x)\pi(y,t)
\end{eqnarray}
where $\epsilon(x,t)=\bar\phi(x,t)-\phi^*(x)$,  $A(x,y)=\delta^2 H[\phi,\pi]/\delta\pi(x)\delta\phi(y)\vert_{\phi=\phi^*,\pi=0}$. The Lyapunov exponents, $\lambda$, of these set of linearized equations are solutions of:
\begin{equation}
det(A+\lambda {\rm I\!I})det(-A+\lambda {\rm I\!I})=0
\end{equation}
We know that $A$ is a negative defined matrix (see eq. (\ref{sta})) for the RD and DD cases. 
Therefore the possible Lyapunov values appear in pairs  $(-\lambda,\lambda)$  which is typical of a Hamiltonian flow. That is, we can define a stable and unstable manifolds  crossing the stationary point $P^*:(\phi^*,0)$. All the trajectories that are solution of eq. (\ref{Heq}) starting from the stationary point pertain to the unstable manifold, $M_u$. This is important from a practical (numerical) point of view if we want to solve the equations of motion: whenever we choose as initial condition $P^*$ we will stay there forever. Therefore, the right strategy is to reconstruct the unstable manifold around $P^*$  and then taking  points $P_0$ pertaining to $M_u$ as initial conditions for solving the equations of motion (see for instance \cite{MG}).

\item The evolution equations (\ref{Heq}) are nonlinear and it could be that for a given $\eta$ there exists, for instance, a couple of pairs $I_1:(\eta,\pi_1)$ and $I_2:(\eta,\pi_2)$  pertaining to the unstable manifold. Therefore both evolve to the stationary state $(\phi^*,0)$ as $t\rightarrow-\infty$ and the quasipotential is:
\begin{equation}
V_0[\eta]=V_0[\phi^*]+min[A[\eta;c_1],A[\eta;c_2]]\quad,\quad A[\eta;c]=\int_c \pi d\phi
\end{equation}
where $c_1$ and $c_2$ are the paths described by the hamiltonian trajectories connecting the stationary state with  $I_1$ and $I_2$ respectively. For the values $\eta_d$ such that $A[\eta_d;c_1]=A[\eta_d;c_2]$ there are  associated two different $\pi_d$ values depending on how we approach $\eta_d$ from $c_1$ or $c_2$. Therefore the quasipotential $V_0$ has a discontinuous first derivative at $\eta_d$. This phenomena is called {\it Lagrangian transition} and it appears only in systems at nonequilibrium stationary states \cite{seminal,Bertini3}. 

\item For the RD and DD cases, there is a family of Langevin equations with a priori known stationary solutions. For instance, the Langevin equation such that:
\begin{eqnarray}
F[\phi;x]&=&-\frac{1}{2}h[\phi;x]^2\frac{\delta V[\phi]}{\delta\phi(x)}\quad\text{(RD case)}\nonumber\\
G[\phi;x]&=&-\frac{1}{2}\chi[\phi;x]\nabla\frac{\delta V[\phi]}{\delta\phi(x)}\quad \text{(DD case)}\label{eq0}
\end{eqnarray}
have the quasipotential $V_0[\phi]=V[\phi]$ for any functionals $h[\phi;x]$, $\chi[\phi;x]$ and $V[\phi]$ such that they meet the conditions about stability (\ref{sta}) and behavior at the boundaries $\delta V[\phi]/\delta\phi(x)=0\,\forall x\in\partial\Lambda$. These particular cases are relevant because permits us to build Langevin equations with an {\it a priori} given stationary state.

\item  The quasi-potential has a relevant dynamic property: It is a Lyapunov functional for $\phi_D(t)$ and $\bar\phi(-t)$. That is, we can proof that if
\begin{equation}
S[\phi]=V_0[\phi]-V_0[\phi^*]
\end{equation}
then
\begin{equation}
\frac{dS[\phi_D(t)]}{dt}\leq 0 \quad\text{and}\quad \frac{dS[\bar\phi(-t)]}{dt}\leq 0
\end{equation}
where $\phi_D(t)=\{\phi_D(x,t), x\in\Lambda\}$ and $\bar\phi(t)=\{\bar\phi(x,t), x\in\Lambda\}$ are the solutions of equations (\ref{det}) and (\ref{Heq}) respectively.
Moreover,
\begin{equation}
\lim_{t\rightarrow\infty}\frac{dS[\phi_D(t)]}{dt}=0\quad\text{and}\quad  \lim_{t\rightarrow\infty}\frac{dS[\bar\phi(-t)]}{dt}=0
\end{equation}
In other words, the time evolution of the deterministic dynamics, $\phi_D(t)$ and the solutions of the so-called  {\it deterministic adjoint dynamics}, $\bar\phi(-t)$, tend to minimize the quasi-potential at all times. 
Let us prove these properties for the RD case. From the definition of $S$ we write:
\begin{equation}
\frac{dS[\phi(t)]}{dt}=\int_{\Lambda}dx\frac{\delta V_0[\phi(t)]}{\delta\phi(x,t)}\partial_t\phi(x,t)
\end{equation}
we know that $\partial_t\phi_D(x,t)=F[\phi_D;x,t]$ and $\partial_t\bar\phi(x,-t)=-F[\bar\phi;x,-t]-h[\bar\phi;x,-t]^2\pi(x,-t)$. Therefore:
\begin{eqnarray}
\frac{dS[\phi_D(t)]}{dt}&=&\int_{\Lambda}dx\frac{\delta V_0[\phi_D(t)]}{\delta\phi_D(x,t)}F[\phi_D;x,t]\nonumber\\
\frac{dS[\bar\phi(-t)]}{dt}&=&\int_{\Lambda}dx\frac{\delta V_0[\bar\phi(-t)]}{\delta\bar\phi(x,-t)}\left(-F[\bar\phi;x,-t]-h[\bar\phi;x,-t]^2\pi(x,-t)\right) 
\end{eqnarray}
we now use the Hamilton-Jacobi equation (\ref{HJ}) and we get the desired result:
\begin{eqnarray}
\frac{dS[\phi_D(t)]}{dt}&=-\frac{1}{2}&\int_{\Lambda}dx h[\phi_D;x,t]^2\left(\frac{\delta V_0[\phi_D(t)]}{\delta\phi_D(x,t)}\right)^2\leq 0\nonumber\\
\frac{dS[\bar\phi(-t)]}{dt}&=-\frac{1}{2}&\int_{\Lambda}dx h[\bar\phi;x,-t]^2\pi(x,-t)^2\leq 0
\end{eqnarray}
Finally the unique state in which such derivatives are equal to zero is the stationary state $\phi^*$. Therefore $S$ is a positive defined functional that decrease  monotonously  with time until it reaches the stationary state. One can do a very similar computation for the DD case.
\item  Some different DD models may have the same quasipotential. Let us assume a model with $G[\phi;x]$ and $\chi[\phi;x]$ having a stationary state $\phi^*$ and a quasipotential $V_0[\phi]$. Then any other model with $\bar G[\phi;x]=G[\phi;x]+S[x]$ and $\chi[\phi;x]$ with $\nabla S(x)=0$ $\forall x\in\Lambda$ has an stationary state $\bar\phi^*=\phi^*$ and a quasipotential $\bar V_0[\phi]=V_0[\phi]$. This doesn't imply that they are describing the same physical situation. For instance their currents $\bar G$ and $G$  are different. Therefore we can conclude with this example that the quasipotential does not always contain all the relevant information about the system stationary state.
\end{itemize}

It is a very difficult task to get explicitly $V_0[\phi]$ from the above definitions. Up to our knowledge it has been obtained  explicitly only in a few  one dimensional models for instance the boundary driven symmetric exclusion process (SSEP) and the Kipnis, Marchioro and Presutti model (KMP)  \cite{Derrida,Bertini1}. In both cases it is found that $V_0$ presents a  non-local structure that it is build through an auxiliary field that it is solution of a non-linear second order differential equation that includes the boundary conditions. These results illuminates the complex mathematical structure of nonequilibrium stationary states. Finally let us mention some serious efforts to define  perturbative schemes around some known exact stationary state (see for instance \cite{Bouchet}). 
 
 \section{Macroscopic Reversibility}
 We have exposed the way to compute  the quasi-potential, $V_0$, from the Langevin equation that defines the system mesoscopic dynamics. At this point, it seems that there is no formal distinction between a system being  in an equilibrium or in a non-equilibrium stationary state.  In any case we have to build $V_0$ from our Hamilton variational principle. Then, from this point of view, there are some fundamental questions that arise:  how can we know whether a system in an equilibrium or in a nonequilibrium stationary state?  We already commented above about the possibility that $V_0$ had some non-analyticities in its domain of definition. That fact contrast with the regular behavior we know from the equilibrium ensemble when looking the corresponding Free Energy Functional. Is therefore, a systematic non-analytic behavior  the main difference between an equilibrium and a nonequilibrium stationary state? Are there any other differences between the two cases?   We could try to create a catalog of $V_0$'s by observing the different mathematical properties that can arise from the equations we explicitly got and then we could discuss about which ones are compatible with an equilibrium state or not. This mathematical approach could be possible but  we think that  trying to characterize equilibrium or nonequilibrium via the structural form of $V_0$ is not the correct approach. From a physical point of view there is a clear cut between the two cases and therefore, the mathematical peculiarities of $V_0$ are a consequence of it.  Therefore we should find an a priori property to know whether or not our system is in an equilibrium state or not.  The key notion here is the behavior of our system under a time reverse operation. That will clarify most of the above questions and comments.
 
 From a macroscopic point of view a system (RD or DD) has two deterministic well defined dynamics: (I) the deterministic evolution equations (\ref{det}) whose solution, $\phi_D(t)$  is the most probable path that the system follows  when it {\it relax} from an arbitrary $\eta$ initial condition to the stationary state $\phi^*$ and (II) the deterministic adjoint dynamics, $\bar\phi(t)$ that is solution of the Hamiltonian equations of motion (\ref{Heq}) and it represents the most probable path that follows a  fluctuation from the stationary state $\phi^*$ to $\eta$. 
We could agree that $\phi_D(t)$ and $bar\phi(-t)$ should differ for a system in a nonequilibrium stationary state because the external mechanisms or boundary conditions that create such state is, for definition, sensitive to the time arrow. Therefore this property seems to be relevant in order to discern between equilibrium and nonequilibrium stationary states. Let us go deeper into it by defining the concept of macroscopic time reversibility and by seeing which are the consequences for a system that has it.
 
 \begin{itemize}
\item{\it Definition:} A system is called {\bf macroscopically time-reversible}  when $\phi_D(x,t)=\bar\phi(x,-t)$. In other words,  the most probable path to create a fluctuation is just the time reversed one to relax the fluctuation using the deterministic dynamic equation. 
\end{itemize}   
 Let us first see the consequences of a system being macroscopically time-reversible.

\subsection{RD case:}

Macroscopic time-reversibility implies in this case that 
\begin{equation}
\bar\phi(x,t)=\phi_D(x,-t)\Rightarrow \partial_t\bar\phi(x,t)=-F[\bar\phi;x,t]\label{mrev}
\end{equation}
At the same time $\bar\phi$ is solution of the eq. (\ref{HJ}):
\begin{eqnarray}
\partial_t\bar\phi(x,t)&=&F[\bar\phi;x,t]+h[\bar\phi;x,t]^2\pi[x,t]\nonumber\\
\partial_t\pi(x,t)&=&-\int_\Lambda dy\,\pi(y,t)\left[\frac{\delta F[\bar\phi;y,t]}{\delta \bar\phi(x,t)}+\frac{1}{2}\frac{\delta h[\bar\phi;y,t]^2}{\delta\bar\phi(x,t)}\pi(y,t) \right]\label{HJRD}
\end{eqnarray}
Equating the first equation from (\ref{HJRD}) with eq.(\ref{mrev}) we get
\begin{equation}
\partial_t\bar\phi(x,t)=F[\bar\phi;x,t]+h[\bar\phi;x,t]^2\pi(x,t)=-F[\bar\phi;x,t]\Rightarrow \pi(x,t)=-\frac{2F[\bar\phi;x,t]}{h[\bar\phi;x,t]^2}\label{r0}
\end{equation}
The $\pi(x,t)$ obtained is also solution of the second of the equations of motion in (\ref{HJRD}). After its substitution we  get:
\begin{equation}
\int_{\Lambda}dy\, F[\phi;y]\left(\frac{\delta}{\delta\phi(y)}\left[\frac{F[\phi;x]}{h[\phi;x]^2}\right]-\frac{\delta}{\delta\phi(x)}\left[\frac{F[\phi;y]}{h[\phi;y]^2}\right]\right)=0 \label{r1}
\end{equation}
This is a necessary condition for our system defined by $F$ and $h$ to be {\it macroscopically time-reversible}.

We also get properties of the associated quasi-potential $V_0$ for these systems if we use the fact that  $\pi(x,t)=\delta V_0[\bar\phi]/\delta\bar\phi(x,t)$. Therefore
\begin{eqnarray}
&&\frac{\delta V_0[\phi]}{\delta\phi(x)}=-\frac{2F[\phi;x]}{h[\phi;x]^2}\label{r3a}\\
&&\int_{\Lambda}dy\, F[\phi;y]\left(\frac{\delta^2V_0[\phi]}{\delta\phi(x)\delta\phi(y)}-\frac{\delta^2V_0[\phi]}{\delta\phi(y)\delta\phi(x)}\right)=0 \label{r3b}
\end{eqnarray}

Let us do some remarks:
\begin{itemize}
\item For macroscopic time reversible systems the first functional derivative of $V_0$ is {\it always} a local functional (whenever $F[\phi;x]$ is local) independently on the structure of $V_0$. Moreover, if the system has some boundary condition then  $F[\phi;x]=0$ $\forall x\in\partial\Lambda$.

\item When 
\begin{equation}
D[\phi;x,y]\equiv\frac{\delta}{\delta\phi(y)}\left(\frac{F[\phi;x]}{h[\phi;x]^2}\right)\label{c1}
\end{equation}
 is  symmetric: $D[\phi;x,y]=D[\phi;y,x]$, then the condition (\ref{r1}) is fulfilled and the system is macroscopically time reversible. In this case $V_0$ is continuous, with continuous first and second functional derivatives. Moreover $V_0[\phi]$ can be obtained directly by integrating eq.(\ref{r3a}).  Let us remind here that in this paper we are considering situations with one unique stationary state. In case of several stationary states the theory should be generalized when constructing $V_0$  (see for instance Ref.\cite{seminal}) and one may have some non-analytic behavior on $V_0$  even in equilibrium when, for instance, there are coexisting phases.

\item When $D[\phi;x,y]\neq D[\phi;y,x]$ the condition given by eq.(\ref{r1}) typically fails (in any case one should check that the integral is not zero for any field). Therefore the system is not macroscopically time reversible and, as we will see below, it has a nonequilibrium stationary state.

\item A RD dynamics that is built from a  twice differential $V[\phi]$ functional, a noise intensity $h$ (see eq.(\ref{eq0})) and with boundary conditions (if any) such that $F[\phi;x]=0 \quad \forall x\in\partial\Lambda$ is macroscopically time-reversible and $V_0[\phi]=V[\phi]$. A typical example of these potentials are the ones of the form:
\begin{equation}
V_0[\phi]=\int_\Lambda dx \,v[\phi;x]\label{eqGL}
\end{equation}
with $v[\phi;x]$ having the property 
\begin{equation}
\frac{\delta^2 v[\phi;x]}{\delta\phi(v)\delta\phi(z)}=\frac{\delta^2 v[\phi;x]}{\delta\phi(z)\delta\phi(v)} \quad\forall\quad v,z\in\Lambda
\end{equation}
then
\begin{equation}
F[\phi;x]=-\frac{1}{2}h[\phi;x]^2\int_{\Lambda}dy \frac{\delta v[\phi;y]}{\delta \phi(x)}
\end{equation}
For instance if we choose the Ginzburg-Landau form:
\begin{equation}
v[\phi;x]=\frac{1}{2}(\nabla\phi)^2+w(\phi(x))\label{GL1}
\end{equation}
with $w(\lambda)$ just any one dimensional function. Then we find 
\begin{equation}
F[\phi;x]=\frac{1}{2}h[\phi;x]^2\left(\Delta\phi(x)-\frac{dv(\lambda)}{d\lambda}\biggr\vert_{\lambda=\phi(x)}\right)
\end{equation}
The corresponding Langevin dynamics is the well known Hohenberg-Halpering model A \cite{Hoh}. 
\end{itemize}
\subsection{DD case:}

A DD system is {\it macroscopically time reversible} when the most probable path that is solution of the Hamilton equations (\ref{HJ})
\begin{eqnarray}
\partial_t\bar\phi(x,t)&=&-\nabla\cdot G[\bar\phi;x,t]-\nabla\left(\chi[\bar\phi;x,t]\nabla\pi(x,t)\right)\nonumber\\
\partial_t\pi(x,t)&=&-\int_{\Lambda}dy\nabla\pi(y,t)\cdot\left[\frac{\delta G[\bar\phi;y,t]}{\delta\bar\phi(x,t)}+\frac{1}{2}\frac{\delta\chi[\bar\phi;y,t]}{\delta\bar\phi(x,t)}\nabla\pi(y,t)\right]\label{HJdd}
\end{eqnarray}
with $\bar\phi(x,-\infty)=\phi(x)^*$ and $\pi(x,-\infty)=0$, is also solution of the time reversed deterministic equation:
\begin{equation}
\partial\bar\phi(x,t)=\nabla\cdot G[\bar\phi;x,t]
\end{equation}
Therefore and similarly to the RD case we get two necessary conditions over $G$ and $\chi$
\begin{equation}
\bar G[\phi;x]\equiv G[\phi;x]-S(x)=-\frac{1}{2}\chi\cdot\nabla\frac{\delta V_0[\phi]}{\delta\phi(x)}\quad,\quad \nabla\cdot S[x]=0\label{d0}
\end{equation}
where $V_0$ is independent on $S(x)$ as we commented above 
and
\begin{equation}
\int_\Lambda dy\,\sum_\gamma {\mathcal D}_\gamma[\phi;y]\left[\left(\partial_{y,\gamma}\frac{\delta}{\delta\phi(y)}\right){\mathcal D}_\alpha[\phi;x]-\left(\partial_{x,\alpha}\frac{\delta}{\delta\phi(x)}\right){\mathcal D}_\gamma[\phi;y] \right]=0\quad\forall\alpha\label{d1}
\end{equation}
where
\begin{equation}
{\mathcal D}_\alpha[\bar\phi;x]=\sum_{\beta}\chi_{\alpha\beta}^{-1}[\bar\phi;x]\bar G_\beta[\bar\phi;x]
\end{equation}
These equations are very similar to the the ones we obtained for the RD case (see eqs.(\ref{r0},\ref{r1})) and most of the comments there apply here:
\begin{itemize}
\item  The conditions about macroscopic time reversibility are defined  on $\bar G[\phi;x]$. Therefore  $G[\phi;x]+S(x)$ with $\nabla S(x)=0$ is macroscopically time reversible  whenever $G[\phi;x]$ is.  Therefore we may always build DD models with an arbitrary net current that are macroscopic time reversible.

\item For macroscopic time reversible systems $\nabla \delta V_0/\delta\phi(x)$ is {\it always} a local functional (whenever $G[\phi;x]$ is local) independently on the structure of $V_0$. 

\item Whenever 
\begin{equation}
E[\phi;x,y;\alpha,\gamma]=\left(\partial_{x,\alpha}\frac{\delta}{\delta\phi(x)}\right) D_\gamma[\phi;y]\label{c2}
\end{equation}
is symmetric: $E[\phi;x,y;\alpha,\gamma]=E[\phi;y,x;\gamma,\alpha]$, then the condition (\ref{d1}) is met and
the system is macroscopically time reversible. Moreover $V_0$ can be found by integrating eq. (\ref{d0}).

\item
The DD dynamics that is built from a  twice differential $V[\phi]$ functional, a noise intensity $\chi$ (see eq.(\ref{eq0})) and with the appropriate boundary conditions : $G[\phi;x]=0 \quad\forall x\in\partial\Lambda$ is macroscopically time-reversible.
 In particular, if we choose $V_0[\phi]$ of the form (\ref{eqGL}) with $v[\phi;x]$ given by eq. (\ref{GL1}), we get:
\begin{equation}
G_\alpha[\phi;x]=\frac{1}{2}\sum_{\beta=1}^d\chi_{\alpha\beta}[\phi;x]\partial_\beta\left(\Delta\phi(x)-\frac{dv(\lambda)}{d\lambda}\biggr\vert_{\lambda=\phi(x)}\right)
\end{equation}
This expression corresponds to the Hohenberg-Halpering model B \cite{Hoh}.
\end{itemize}

 \section{The Fundamental Principle: The adjoint dynamics}
 
Bertini and co-workers defined the adjoint dynamics by extending large deviation properties of several microscopic stochastic models to diffusive mesoscopic systems \cite{Bertini}. 
In fact they generalize the Einstein's proposal about fluctuations of systems at equilibrium in which he connected the probability of having a fluctuation with the minimum reversible work necessary to create it. Their idea is to assume that the probability of any path from an initial stationary state  is equal to the probability of the {\it time reversed} path. They call this property {\it the Fundamental Principle} and it is the definition of the adjoint dynamics that it is just the dynamics associated to the time reversed system.  
 Moreover, they assumed that the adjoint dynamics so defined follows a mesoscopic equation similar, in structure, to the original one. This permits them to obtain the equations of motion for the fields under the action of the adjoint dynamics. Therefore they could compare both set of equations of motion to look for properties due to time symmetries. In fact they saw that this important concept contains the essential ingredients to discriminate between systems at non-equilibrium stationary states from systems at equilibrium. We follow these clever proposals (in our opinion) by Bertini and co-workers \cite{Bertini}  by assuming that the Fundamental Principle is valid to our models. From it we can get the adjoint dynamics for the RD and DD cases. 
 
\subsection{RD's adjoint dynamics:}

Let us define the joint probability of a given path from $t_0$ to $t_1$ knowing that $\phi[t_0]$ is chosen from the stationary distribution:
\begin{equation}
P(\{\phi\}[t_0,t_1]\vert\phi[t_0])=P_{st}[\phi[t_0]]P[\{\phi\}[t_0,t_1]]
\end{equation}

Let us fix a path $\{\phi\}[t_0,t_1]$ and its time reversed image: $\{\tilde\phi\}[-t_1,-t_0]$ where $\tilde\phi(x,t)=\phi(x,-t)$. The {\it Fundamental  Principle} states that 
\begin{equation}
P(\{\phi\}[t_0,t_1]\vert\phi[t_0])=P^*(\{\tilde\phi\}[-t_1,-t_0]\vert\tilde\phi[-t_1])\label{FPrin}
\end{equation}
where $P^*$ is the probability of a path for the adjoint dynamics.
We assume now that $\Omega\rightarrow\infty$ and using eqs. (\ref{pathg}), (\ref{hamRD}) and (\ref{v0}) we get
\begin{equation}
P(\{\phi\}[t_0,t_1]\vert\phi[t_0])\simeq\exp\left[-\Omega R[\{\phi\}[t_0,t_1]]\right]\label{FP1}
\end{equation} 
where
\begin{eqnarray}
R[\{\phi\}[t_0,t_1]]&=& V_0[\phi(t_0)]+\inf_\pi\int_{t_0}^{t_1}dt\left[\int_\Lambda dx\, \pi(x,t)\partial_t\phi(x,t)-H[\phi(t),\pi(t)]\right]\nonumber\\
&=& V_0[\phi(t_0)]+\int_{t_0}^{t_1}dt\int_\Lambda dx \left(\frac{\partial_t\phi(x,t)-F[\phi;x,t]}{2 h[\phi;x,t]}\right)^2
 \end{eqnarray}
 Then, the Fundamental Principle implies, in this limit,
 \begin{eqnarray}
&&V_0[\phi[t_0]]+\frac{1}{2}\int_{t_0}^{t_1}dt\int_\Lambda dx \left(\frac{\partial_t\phi(x,t)-F[\phi;x,t]}{h[\phi;x,t]}\right)^2\nonumber\\
&=&V_0[\phi[t_1]]+\frac{1}{2}\int_{t_0}^{t_1}dt\int_\Lambda dx \left(\frac{\partial_t\phi(x,t)+F^*[\phi;x,t]}{h^*[\phi;x,t]}\right)^2\label{rel}
\end{eqnarray}
 for any path $\{\phi\}[t_0,t_1]$. Where we have assumed that the adjoint dynamics is defined by the Langevin equation:
 \begin{equation}
\partial_t\tilde\phi(x,t)=F^*[\tilde\phi;x,t]+h^*[\tilde\phi;x,t]\xi(x,t)\label{emotad}
\end{equation}
 with $\xi$ being an uncorrelated white noise.
 
Let us show that equation (\ref{rel}) defines the mathematical form of $F^*$ and $h^*$ functionals. Let us assume that $V_0$ is differentiable along the path chosen. Then
\begin{equation}
V_0[\phi[t_1]]-V_0[\phi[t_0]]=\int_{t_0}^{t_1}dt\,\partial_tV_0[\phi[t]]=\int_{t_0}^{t_1}dt\int_\Lambda dx\frac{\delta V_0[\phi[t]]}{\delta\phi(x,t)}\partial_t\phi(x,t)\label{tr}
\end{equation}
therefore equation (\ref{rel}) can be written:
\begin{eqnarray}
&&\int_{t_0}^{t_1}dt\int_\Lambda dx\,\biggl\{\frac{1}{2}\left(\frac{1}{h[\phi;x,t]^2}-\frac{1}{h^*[\phi;x,t]^2}\right)(\partial_t\phi(x,t))^2\nonumber\\
&&-\left(\frac{F^*[\phi;x,t]}{h^*[\phi;x,t]^2}+\frac{F[\phi;x,t]}{h[\phi;x,t]^2}+\frac{\delta V_0[\phi[t]]}{\delta\phi(x,t)}\right)\partial_t\phi(x,t)\nonumber\\
&&-\frac{F^*[\phi;x,t]}{2h^*[\phi;x,t]^2}+\frac{F[\phi;x,t]^2}{2h[\phi;x,t]^2}
\biggr\}=0
\end{eqnarray}
for any path and any time interval. Then, we can fix a time $t$,  and change paths such that they have arbitrary values for $\partial_t\phi(x,t)$. Therefore the coefficients of the time derivatives should be equal to zero. Thus:
\begin{eqnarray}
(\partial_t\phi(x,t))^2&:&h^*[\phi;x,t]=h[\phi;x,t]\nonumber\\
(\partial_t\phi(x,t))^1&:&F^*[\phi;x,t]=-F[\phi;x,t]-h[\phi;x,t]^2\frac{\delta V_0[\phi]}{\delta\phi(x,t)}\nonumber\\
(\partial_t\phi(x,t))^0&:&\int_\Lambda dx \frac{F^*[\phi;x,t]^2-F[\phi;x,t]^2}{h[\phi;x,t]^2}=0
\end{eqnarray}
The first equation indicates that the adjoint dynamics has the same noise intensity as the direct dynamics. The second one shows that its  deterministic part is different and it depends on the quasipotential. Finally, it can be easily shown that the last equation is just the Hamilton-Jacobi equation (\ref{HJ}).

\subsection{DD's adjoint dynamics:}

In this case the Fundamental Principle should be applied to path probabilities for the variables that define the state of the system: $(\phi,j)$:
\begin{equation}
P(\{\phi,j\}[t_0,t_1]\vert\phi[t_0])=P^*(\{\tilde\phi,\tilde \j\}[-t_1,-t_0]\vert\tilde\phi[-t_1])\label{FPrin2}
\end{equation}
where
\begin{equation}
P(\{\phi,j\}[t_0,t_1]\vert\phi[t_0])=P_{st}[\phi[t_0]]P[\{\phi,j\}[t_0,t_1]]
\end{equation}
$P[\{\phi,j\}[t_0,t_1]]$ is defined in eq. (\ref{path2}) and the fields associated to the time reversed path are such: $\tilde\phi(x,t)=\phi(x,-t)$, $\tilde \j(x,t)=-j(x,-t)$.
Again, when $\Omega\rightarrow\infty$ we get
\begin{eqnarray}
V[\phi(t_1)]-V[\phi(t_0)]&=&\frac{1}{2}\int_{t_0}^{t_1}dt\,\int_\Lambda dx\,\biggl[\left(j(x,t)-G[\phi;x,t]\right)\cdot\chi^{-1}[\phi;x,t]\left(j(x,t)-G[\phi;x,t]\right)\nonumber\\
&-&\left(j(x,t)+G^*[\phi;x,t]\right)\cdot{\chi^*}^{-1}[\phi;x,t]\left(j(x,t)+G^*[\phi;x,t]\right)\label{DDrel}
\biggr]
\end{eqnarray}
where we have assumed that the adjoint dynamics is defined by the Langevin equation
\begin{equation}
\partial_t\tilde\phi(x,t)+\nabla\cdot \tilde\j(x,t)=0
\end{equation}
with the constitutive equation:
\begin{equation}
\tilde\j_{\alpha}(x,t)=G^*_{\alpha}[\tilde\phi;x,t]+\sum_{\beta=1}^d\sigma^*_{\alpha,\beta}[\tilde\phi;x,t]\psi_{\beta}(x,t)  \quad \alpha=1,\ldots,d\label{emotad2}
\end{equation}
 and 
 \begin{equation}
\chi^*_{\alpha,\beta}[\tilde\phi;x,t]=\sum_{\gamma=1}^d\sigma^*_{\alpha,\gamma}[\tilde\phi;x,t]\sigma^*_{\beta,\gamma}[\tilde\phi;x,t]
\end{equation}
 Relation (\ref{DDrel}) should hold for any path and therefore for any value of $j(x,t)$ at any time. Then
\begin{eqnarray}
(j(x,t))^2&:&\chi^*[\phi;x,t]=\chi[\phi;x,t]\nonumber\\
(j(x,t))^1&:&G^*[\phi;x,t]=-G[\phi;x,t]-\chi[\phi;x,t]\nabla\frac{\delta V_0[\phi]}{\delta\phi(x,t)}\nonumber\\
(j(x,t))^0&:&H[\phi,\frac{\delta V_0[\phi]}{\delta\phi(x,t)}]=0
\end{eqnarray} 
where $H[\phi,\pi]$ is given by eq.(\ref{hamDD}). 

There are two observations to be done:
\begin{itemize}
\item By construction, the quasipotential associated to the adjoint dynamics is the same that for the original one: $V_0[\phi]$.
\item The equation of motion for the deterministic part of the adjoint dynamics (eq.(\ref{emotad}) for RD or eq.(\ref{emotad2}) for DD without the noise term) is equal to the time reversed equation of motion that define the most probable path to create a fluctuation (eq.(\ref{Heq})).  Therefore: $\tilde\phi_D(x,t)=\bar\phi(x,-t)$. That is, the deterministic path that is followed by the adjoint dynamics to relax a initial configuration $\phi_0$ to the stationary state $\phi^*$  is just the time reversed most probable path that goes from the stationary state $\phi^*$ to $\phi_0$ by the normal deterministic dynamics. 
\end{itemize}

\subsection{Equilibrium vs non-equilibrium}

We know that the definition of the equilibrium concept  is a very subtle issue.  Nevertheless we could probably agree that at the macroscopic level  equilibrium is a state where the macroscopic properties of a system do not change in time and where there are no flux of any type across it. Moreover, at the mesoscopic level we could use the Gibbs's definition of equilibrium: the state where the probability density function for a configuration is a time-independent solution of Liouville’s equation. The definition of equilibrium in the MFT context is  difficult because we are dealing with mesoscopic systems with a priori no connection with any underlying mechanical microscopic model.
We get some hint from  the theory by Onsager and Machlup  about fluctuations and relaxation towards the equilibrium of mesoscopic variables \cite{Onsager}. There they assumed that the underlying time reversibility of the microscopic equations of motion should appear in the mesoscopic equations in order to derive their properties near the equilibrium state. Therefore  the time reversibility of the mesoscopic dynamics should be  the essential item that characterizes a system in an equilibrium state. We already know that the {\it  Macroscopic time reversibility} property does not discriminate with enough precision equilibrium from non-equilibrium stationary states.  For instance in a torus we may always have systems being macroscopic time reversible with non zero average currents. That is typical of  nonequilibrium stationary states (see for instance the explicit example given by Bertini et al. in \cite{Bertini2}).  In constrat, the Fundamental Principle above defined contains detailed information on the time-reversed mesoscopic dynamics (adjoint dynamics) and it is the natural place to  understand the consequences of time reversal on the RD and DD models (see eqs. (\ref{FPrin}) or (\ref{FPrin2})). 

We assume that a system at equilibrium should behave identically under a time reversal operation. That is, it should follow the same dynamics independently on the arrow's time. That happens  when the system dynamics follows this proposition: 
\begin{itemize}
\item{{\it Proposition:}} A system is at {\bf equilibrium} when the adjoint dynamics is equal to the original one. 
\end{itemize}
This definition of equilibrium applied to the RD and DD systems implies that $F$ and $G$ should be of the form:
\begin{eqnarray}
F[\phi;x]&=&-\frac{1}{2}h[\phi;x]^2\frac{\delta V_0[\phi]}{\delta\phi(x)}\quad\text{(RD case)}\nonumber\\
G[\phi;x]&=&-\frac{1}{2}\chi[\phi;x]\nabla\frac{\delta V_0[\phi]}{\delta\phi(x)}\quad \text{(DD case)}\label{eq00}
\end{eqnarray}
where $V_0[\phi]$ is solution of the Hamilton-Jacobi equation $H[\phi,\delta V_0/\delta\phi]=0$. Therefore $F$ and $G$ have too the differential properties that we found for the macroscopically time reversible systems:  $D[\phi;x,y]=D[\phi;y,x]$  for the RD and $E[\phi;x,y;\alpha,\gamma]=E[\phi;y,x;\gamma,\alpha]$ for DD, where $D$ is given in (\ref{c1}) and $E$ in  (\ref{c2}).

We point out some comments:
\begin{itemize}
\item In general:
\begin{equation}
\text{Stationary State is Equilibrium}\Rightarrow \text{Macroscopic Time Reversible System}
\end{equation}
For systems with RD  the reverse implication is true and therefore having an equilibrium stationary state is equivalent to be Macroscopically Time Reversible. For  systems with DD  the reverse implication isn't true because  only the Macroscopically Time Reversible systems with $S(x)=0$ have an Equilibrium stationary state. 

\item  The deterministic current for systems with DD at equilibrium is zero: $G[\phi^*;x]=0$. It is proven (see for instance Ref.\cite{Bertini}) that for boundary driven diffusive systems the reverse statement  is true and zero deterministic current at the stationary state implies equilibrium. However that is not true in general. There are systems with zero deterministic current at the stationary state and generic power law decay of correlations that is typical of nonequilibrium stationary states \cite{Gar}. 

\end{itemize}

  \section{Correlations}

Another form to get some insight on $V_0$ is the study of the stationary correlations.
There is a direct relation between the quasipotential around the stationary state and the correlations. Let us define 
\begin{equation}
Z_0=\int D\phi P_{st}[\phi]\simeq\int D\phi \exp\left[-\Omega V_0[\phi]\right] \quad , \Omega\rightarrow\infty\label{star0}
\end{equation}
It can be shown that the two body correlations on the stationary state is given by
 \begin{equation}
C_2(x,y)\equiv\lim_{\Omega\rightarrow\infty}\Omega\langle(\phi(x)-\phi^*(x))(\phi(y)-\phi^*(y))\rangle_{st}=-2\frac{\delta\log \bar Z_0[V_2]}{\delta V_2(x,y)}\label{eqcor0}
\end{equation}
where
\begin{equation}
\bar Z_0[V_2]=\int D\omega\exp\left[-\frac{1}{2}\int_\Lambda dxdy\,V_2(x,y)\omega(x)\omega(y) \right]
\end{equation}
and
\begin{equation}
V_2(x,y)=\frac{\delta^2 V_0[\phi]}{\delta\phi(x)\delta\phi(y)}\biggr\vert_{\phi=\phi^*}\label{V2}
\end{equation}
Where we have assumed that $V_0$ can be expanded around its stationary state $\phi^*$ and we have done the change of variables $\omega(x)=\sqrt\Omega(\phi(x)-\phi^*(x))$ in eq.(\ref{star0}).

$\bar Z_0$ can be explicitly computed because is a gaussian-like integral:
\begin{equation}
\bar Z_0[V_2]=cte (\det(V_2))^{-1/2}
\end{equation}
and after substituting in eq.(\ref{eqcor0}) finally we get
\begin{equation}
C_2(x,y)=V_2^{-1}(x,y) \label{eqcor}
\end{equation}
We can obtain the same result by constructing the quasipotential (\ref{action2}) by solving the Hamilton evolution equations (\ref{Heq}) near the stationary state (see Appendix IV).
That is, the two body correlations are directly related with the curvature of the quasipotential around the stationary state and, apart of their own relevance as measured observable that can be checked by experimentalist, it helps us to understand the mathematical structure of the quasipotential. For instance if correlations are long range that would indicate that the relations (\ref{eq00}) do not apply  because they imply local behavior of correlations. 

 From now on, we are going to focus  into the stationary two body correlations.
In order to find a way to compute $C_2$ in a self-consistent way we need to go through a more elaborate scheme \cite{Gardiner,Bertini,Bouchet}.

Let us define first the {\it Generating Functional}:
\begin{equation}
Z[b]=Z[0]\int D\phi\, P_{st}[\phi]\exp\left[\Omega\int_\Lambda dx\, b(x)\phi(x)\right]\label{GF1}
\end{equation}
where $b(x)$ is a kind of auxiliary external field. We know from this expression that the $n$-body correlations at the stationary state (at zero field) is given by
\begin{equation}
\langle\phi(x_1)\ldots\phi(x_n)\rangle_{st}=\frac{1}{\Omega^nZ[0]}\frac{\delta^nZ[b]}{\delta b(x_1)\ldots\delta b(x_n)}\biggr\vert_{b=0}
\end{equation}
The truncated n-body correlations are defined by:
\begin{equation}
\langle\phi(x_1)\ldots\phi(x_n)\rangle_{st}^c=\frac{1}{\Omega^n}\frac{\delta^nW[b]}{\delta b(x_1)\ldots\delta b(x_n)}\biggr\vert_{b=0}
\end{equation}
where $W[b]=\ln Z[b]$,
\begin{equation}
\langle\phi(x_1)\ldots\phi(x_n)\rangle_{st}^c=\langle(\phi(x_1)-\langle\phi(x_1)\rangle_{st})\ldots(\phi(x_n)-\langle\phi(x_n)\rangle_{st}) \rangle_{st} \quad n>1
\end{equation}
and $\langle\phi(x)\rangle_{st}^c=\langle\phi(x)\rangle_{st}$.

We know that 
 $P_{st}[\phi]\propto\exp[-\Omega V_0[\phi]]$ when $\Omega\rightarrow\infty$. Then the Generating functional (\ref{GF1}) can be written:
\begin{equation}
Z[b]\propto\int D\phi \exp[-\Omega\mathcal{F}[\phi,b]]\quad,\quad \mathcal{F}[\phi,b]=V_0[\phi]-\int_{\Lambda}dx\,b(x)\phi(x)\label{star1}
\end{equation}
Let us define $\phi^*[b]$ as the field that minimizes $\mathcal{F}$ and let us assume that $\mathcal{F}[\phi,b]$ is differentiable on $\phi$'s around $\phi^*[b]$, then
\begin{equation}
\mathcal{F}[\phi,b]=\mathcal{F}[\phi^*[b]]+\frac{1}{2}\int_\Lambda dx dy\,\frac{\delta^2\mathcal{F}[\phi,b]}{\delta\phi(x)\phi(y)}\biggr\vert_{\phi=\phi^*[b]}(\phi(x)-\phi^*[x;b])(\phi(y)-\phi^*[y;b])+\ldots
\end{equation}
where $\phi^*[b]$ is solution of
\begin{equation}
\frac{\delta\mathcal{F}[\phi,b]}{\delta\phi(x)}\biggr\vert_{\phi=\phi^*[b]}=0\quad \Rightarrow\quad \frac{\delta V_0[\phi]}{\delta\phi(x)}\biggr\vert_{\phi=\phi^*[b]}=b(x)
\end{equation}
we observe that $\phi^*[0]=\phi^*$, the minimum of $V_0[\phi]$. 

Therefore we get an expansion of the Generating Functional:
\begin{equation}
Z[b]\propto e^{-\Omega\mathcal{F}[\phi^*[b],b]}\int D\omega\exp\left[-\frac{1}{2}\int_\Lambda dxdy \frac{\delta^2V_0[\phi]}{\delta\phi(x)\phi(y)}\biggr\vert_{\phi=\phi^*[b]}\omega(x)\omega(y)+O(\Omega^{-1/2})\right]\label{star2}
\end{equation}
where we  $w(x)=\sqrt{\Omega}(\phi(x)-\phi^*[x;b])$. We see that this expression has some meaning whenever $V_0$ is differentiable and convex around $\phi^*[b]$. Convexity also guarantees that there is a one to one relation between $b$ and $\phi^*[b]$. It may also be shown that  
\begin{equation}
\frac{\delta\mathcal{F}[\phi^*[b],b]}{\delta b(x)}=-\phi^*[x;b]
\end{equation}
That is, $\mathcal{F}[b]\equiv\mathcal{F}[\phi^*[b],b]$ is the Legendre transform of $V_0[\phi]$. We can now relate $\mathcal{F}$ with the correlations:
\begin{equation}
W[b]=-\Omega\mathcal{F}[\phi^*[b]]+O(\Omega^0)
\end{equation}
and
\begin{equation}
\lim_{\Omega\rightarrow\infty}\Omega^{n-1}\langle\phi(x_1)\ldots\phi(x_n)\rangle_{st}^c=-\frac{\delta^n\mathcal{F}[\phi^*[b]]}{\delta b(x_1)\ldots\delta b(x_n)}\biggr\vert_{b=0}\equiv C_n(x_1,\ldots,x_n)
\end{equation}
where $\langle\phi(x)\rangle_{st}^c=\langle\phi(x)\rangle_{st}=\phi^*(x)=\phi^*[x;0]$.

At this point we can use a trick to build a  set of closed equations for the correlation functions. The Hamilton-Jacobi equation applied to $\phi^*(x;b)$ is given by:
\begin{equation}
H\left[\phi^*[b],\frac{\delta V_0[\phi]}{\delta\phi}\biggr\vert_{\phi=\phi^*[b]}\right]=H\left[\phi^*[b],b\right]=0
\end{equation}
where $H[\phi,\pi]$ is given by eqs.(\ref{hamRD}) and (\ref{hamDD}) for the RD and DD respectively. We know that 
\begin{equation}
\phi^*[x;b]=\phi^*(x)+\int_\Lambda dy \,C_2(x,y)b(y)+O(b^2)
\end{equation}
 then we can implement a perturbative expansion on $H$ around $b=0$:
\begin{eqnarray}
&&\int_\Lambda dx\, dy\, b(x)b(y)\biggl[\frac{\delta^2H[\phi,\pi]}{\delta\pi(x)\delta\pi(y)}\biggr\vert_{\substack{\phi=\phi^* \\\pi=0}}+\int_\Lambda dz\biggl[\frac{\delta^2H[\phi,\pi]}{\delta\phi(x)\delta\pi(z)}\biggr\vert_{\substack{\phi=\phi^* \\\pi=0}}C_2(z,y)\nonumber\\
&+&\frac{\delta^2H[\phi,\pi]}{\delta\phi(y)\delta\pi(z)}\biggr\vert_{\substack{\phi=\phi^* \\\pi=0}}C_2(z,x)\biggr]
\biggr]=O(b^3)\quad \forall\, b
\end{eqnarray}
where we have used the fact that $H$ is a second order polynomial in $\pi$ and that $H[\phi^*,0]=0$. Therefore
\begin{eqnarray}
&&\int_\Lambda dz\biggl[\frac{\delta^2H[\phi,\pi]}{\delta\phi(x)\delta\pi(z)}\biggr\vert_{\substack{\phi=\phi^*\\ \pi=0}}C_2(z,y)+\frac{\delta^2H[\phi,\pi]}{\delta\phi(y)\delta\pi(z)}\biggr\vert_{\substack{\phi=\phi^* \\\pi=0}}C_2(z,x)\biggr]\nonumber\\
&=&-\frac{\delta^2H[\phi,\pi]}{\delta\pi(x)\delta\pi(y)}\biggr\vert_{\substack{\phi=\phi^* \\\pi=0}}\label{co}
\end{eqnarray}

Let us explicitly apply eq.(\ref{co}) to the RD and DD cases.

\subsection{RD case:}

We substitute $H$ from eq.(\ref{hamRD}) into (\ref{co}) and we get:
\begin{equation}
\int_\Lambda dz \left[B(x,z)\bar C(z,y)+B(y,z)\bar C(z,x)\right]=-\delta(x-y)\label{twobody}
\end{equation}
where
\begin{equation}
C_2(x,y)=h[\phi^*;x]h[\phi^*;y]\bar C(x,y)
\end{equation}
and
\begin{equation}
B(x,y)=\frac{h[\phi^*;y]}{h[\phi^*;x]}\frac{\delta F[\phi;x]}{\delta\phi(y)}\biggr\vert_{\phi=\phi^*}\label{dB}
\end{equation}
Observe that $B$ maybe non-symmetric on its arguments while $\bar C$ is symmetric by construction. We can think this equation as a representation of the linear operator equation:
\begin{equation}
B\bar C+\bar C B=-I\label{eqcorr}
\end{equation}
with $I$ the identity operator. The formal solution can be found by using the fact that $\partial/\partial\alpha e^{\alpha B}=Be^{\alpha B}$. Then
\begin{equation}
\frac{\partial}{\partial\alpha}\left[e^{\alpha B}\bar Ce^{\alpha B^T}\right]=-e^{\alpha B}e^{\alpha B^T} \Rightarrow\bar C=\int_{0}^{\infty}d\alpha\, e^{\alpha B}e^{\alpha B^T}\label{sol}
\end{equation}
where we have assumed that $B$ is negative defined.  A simple representation of this equation can be obtained in the case that $B$ is {\it diagonalizable}, in other words, when  we can apply to $B$ some spectral theorem.  Let 
$v(x;\lambda_n)$ and $w(x;\lambda_n)$ be the set of {\it right} and {\it left} eigenvectors  of $B$ with eigenvalues $\lambda_n$ and $\lambda_n^*$ (complex conjugate of $\lambda_n$) respectively:
\begin{eqnarray}
\int_\Lambda dy\,B(x,y)v(y;\lambda_n)&=&\lambda_n v(x;\lambda_n)\nonumber\\
\int_\Lambda dy\,B(y,x)w(y;\lambda_n)&=&\lambda_n^* w(x;\lambda_n)
\end{eqnarray} 
 The eigenvalues may have real or complex values but because $B$ is  real valued  they appear in pairs when they are complex: $(\lambda,v(x;\lambda)),(\lambda^*,v(x;\lambda^*)=v(x;\lambda)^*)$. We assume that each set form a complete basis on the functional space and that they follow the orthogonality conditions:
 \begin{eqnarray}
 \int_\Lambda dx\, w(x;\lambda_n)^*v(x;\lambda_m)&=&\delta_{n,m}\nonumber\\
 \sum_n w(x;\lambda_n)^*v(y;\lambda_n)&=&\delta(x-y)
 \end{eqnarray}
Finally, the solution (\ref{sol}) can be written:
\begin{equation}
\bar C(x,y)=-\sum_{n,m}\frac{v(x;\lambda_n)v(y;\lambda_m)}{\lambda_n+\lambda_m}\int_\Lambda dz \,\bar w(z;\lambda_n)\bar w(z;\lambda_m)\label{corr}
\end{equation}
We see that the solution is symmetric, $\bar C(x,y)=\bar C(y,x)$,  and real, $\bar C(x,y)^*=\bar C(x,y)$, due to the pairing property of the eigenvalues. 

The solution (\ref{corr}) can be further simplified if $B$ is symmetric: $B(x,y)=B(y,x)$. In this case the right and left eigenvalues and eigenvectors coincide, all of them are real and the eigenvectors form an orthonormal base on the functional space. Therefore
\begin{equation}
\bar C(x,y)=-\frac{1}{2}\sum_n\frac{1}{\lambda_n}v(x;\lambda_n)v(y;\lambda_n)=-\frac{1}{2}B^{-1}(x,y)
\end{equation}
where
\begin{equation}
\int_\Lambda dz B(x,z)B^{-1}(z,y)=\delta(x-y)
\end{equation}

As an example let us apply these results to small deviations from equilibrium. Let us assume that a system at equilibrium has a given quasi-potential $V_0[\phi]$ and its Langevin dynamics is defined by the pair $(F,h)$ where we know that  $F$ is of the form (\ref{eq00}).  Therefore the correlations are given by eq. (\ref{eqcor}): $C_2(x,y)=V_2^{-1}(x,y)$.
Such reference system is now forced to slightly deviate from the equilibrium by imposing a perturbation in the $F$ term. That is, the  Langevin equation for the new system is composed by the pair $(\tilde F,h)$ where
\begin{equation}
\tilde F[\phi;x]=-\frac{1}{2}h[\phi;x]^2\left[1+\epsilon g[\phi;x]\right]\frac{\delta V_0[\phi]}{\delta\phi(x)}\label{new}
\end{equation}
$V_0[\phi]$, $g[\phi;x]$ and $h[\phi;x]$ are given functionals and $\epsilon$ can be used as a perturbative parameter. This change does not modify the system stationary solution $\phi^*$ for any $\epsilon$: $\tilde F[\phi^*;x]=0\quad\forall\epsilon$. Therefore  the quasipotential $V_\epsilon[\phi]$ associated to the dynamics (\ref{new}) has always the same extremal state: $\delta V_\epsilon[\phi]/\delta\phi(x)\vert_{\phi=\phi^*}=0$ for any $\epsilon$ value. 

To compute the correlations for the dynamics defined in eq.(\ref{new}) we need to construct the matrix $B$ given by eq. (\ref{dB}):
\begin{equation}
B(x,y)=(1+\epsilon g[\phi^*;x])\tilde B(x,y)\quad,\quad \tilde B(x,y)=-\frac{1}{2}h[\phi^*;x]h[\phi^*;y]\frac{\delta^2 \tilde V[\phi]}{\delta\phi(x)\delta\phi(y)}\biggr\vert_{\phi=\phi^*}
\end{equation}
and the equation for the correlations (\ref{eqcorr}) can be written as:
\begin{equation}
\tilde G\tilde B\bar C+\bar C \tilde G\tilde B=-I\label{pertur}
\end{equation}
where $\tilde G(x,y)=(1+\epsilon g[\phi;x])\delta(x-y)$. We look for perturbative solutions of this equation:
\begin{equation}
\bar C=\sum_{n=0}^\infty \epsilon^n\bar C_n
\end{equation}
After substituting the last expression into eq.(\ref{pertur}) we obtain order by order in $\epsilon$ the following hierarchy of equations:
\begin{eqnarray}
\tilde B \bar C_0+\bar C_0\tilde B&=&-I\nonumber\\
\tilde B \bar C_n+\bar C_n\tilde B&=&-G\tilde B\bar C_{n-1}-\bar C_{n-1}\tilde B G\quad n>0
\end{eqnarray}
where $\tilde G=I+\epsilon G$ and $G(x,y)=g[\phi;x]\delta(x-y)$. The solutions are:
\begin{eqnarray}
\bar C_0&=&-\frac{1}{2}\tilde B^{-1}\nonumber\\
\bar C_n&=&\int_{0}^{\infty}d\alpha\, e^{\alpha\tilde B}\left(G\tilde B\bar C_{n-1}+\bar C_{n-1}\tilde B G\right)e^{\alpha\tilde B}\quad n>0
\end{eqnarray}
and, in particular,
\begin{equation}
\bar C_1=-\int_{0}^{\infty}d\alpha\, e^{\alpha\tilde B}G e^{\alpha\tilde B}=QAQ^T
\end{equation}
where $Q$ is the matrix that diagonalizes $\tilde B$: $\tilde B=QDQ^T$, that is $Q_{ij}=v_i(\lambda_j)$ where $v(\lambda)$ is the eigenfuntion of $\tilde B$ with eigenvalue $\lambda$ (all in a formal discrete notation) and 
\begin{equation}
A_{i,j}=\frac{(Q^TGQ)_{ij}}{\lambda_i+\lambda_j}
\end{equation}
Observe that $\tilde B$ is by construction a local functional. However its eigenfunctions (that depend on the boundary conditions and the form of $V_0$) may be non local. Therefore the first correction to the correlations could be already quite singular.
We can get more corrections $\bar C_n$ in the same spirit and we could study some general properties of $\bar C$ depending on the $G$ and $V_0$. However this is beyond the scope of this paper. We just wanted to show the possibility to introduce a perturbative scheme and to show the non-trivial changes that may appear in the behavior of the two body correlations.

\subsection{DD case:}
We substitute $H$ from eq.(\ref{hamDD}) into (\ref{co}) and we get:
\begin{equation}
\int_\Lambda dz\,\left[K(x,z)C_2(z,y)+K(y,z)C_2(z,x)\right]=-\sum_{i,j}\frac{\partial}{\partial x_i}\frac{\partial}{\partial y_j}\left[\chi_{ij}[\phi^*;x]\delta(x-y)\right]\label{DDcor}
\end{equation}
where
\begin{equation}
K(x,y)=\frac{\delta}{\delta\phi(y)}\left(-\nabla\cdot G[\phi;x]\right)\biggr\vert_{\phi=\phi^*}
\end{equation}
with $\phi^*$ solution of $\nabla\cdot G[\phi^*;x]=0$.

In the DD case many models are designed starting from a dynamics acting on the system's bulk together with  boundary conditions that drive the system to an equilibrium state. Afterwards just by changing the boundary conditions one manage to introduce to the system flows of energy, mass,... in such a way that the system is driven  into a nonequilibrium stationary state.  Other nonequilibrium stationary states are just due to a bulk dynamical mechanism that introduces directly some current on the system and/or breaks some symmetry on the system.  Moreover, DD permits models with a strict conserved quantity, the average density, that introduces a kind of long range interaction that affects to the system correlations. Let us comment these particular scenarios for the DD correlations. 

\begin{itemize}
\item {\bf Nonequilibrium stationary states driven by boundary conditions:} Let us assume first that the stationary state of our system is the equilibrium one with a given $V_0[\phi]$ for an appropiate set of boundary conditions. The bulk dynamics is defined by $(G,\chi)$ with 
\begin{equation}
G_i[\phi;x]=-\frac{1}{2}\sum_{j}\chi_{ij}[\phi;x]\partial_j\frac{\delta V_0[\phi]}{\delta\phi(x)}
\end{equation}
We know that the corresponding two-body correlation are:
\begin{equation}
C_2^{eq}(x,y)=V_2^{-1}(x,y;\phi_{eq}^*)\quad ,\quad V_2(x,y;\phi)=\frac{\delta^2V_0[\phi]}{\delta\phi(x)\delta\phi(y)}
\end{equation}
with $\phi_{eq}^*$ solution of $G[\phi_{eq}^*;x]=0$ (the current is equal to zero).  If we change the boundary conditions  the system develops non zero currents and therefore we have a nonequilibrium stationary state. Assume that the stationary state $\phi^*$ is given now by the solution of the equation $G[\phi^*;x]=J$ where $J$ is a constant vector that is fixed by the boundary conditions. In this constext it is convenient to decompose the two body correlation function into two terms: 
\begin{equation}
C_2(x,y)=C_2^{leq}(x,y)+C_D(x,y)\quad ,\quad C_2^{leq}(x,y)=V_2^{-1}(x,y;\phi^*) \label{leq}
\end{equation}
The first term is the {\it local equilibrium correlation} and it is  the equilibrium correlation evaluated with the local values of the field $\phi^*$ as if the system were at equilibrium at $x$ with $\phi_{eq}=\phi^*(x)$.  By other hand we see that $C_D=0$ when $J=0$ by construction.  That is,  $C_D(x,y)$, describes in some sense the far from equilibrium part of the correlations.  When we substitute eq.(\ref{leq}) into (\ref{DDcor}) we get the closed equation for $C_D$:
\begin{eqnarray}
&&\sum_i\frac{\partial}{\partial x_i}\left[\alpha_i[\phi^*;x]C_D(x,y)+\frac{1}{2}\sum_j \chi_{ij}[\phi^*;x]\frac{\partial}{\partial x_j}\int_\Lambda dz\,C_2^{leq,-1}(x,z)C_D(z,y)\right]\nonumber\\
&&\sum_i\frac{\partial}{\partial y_i}\left[\alpha_i[\phi^*;y]C_D(x,y)+\frac{1}{2}\sum_j \chi_{ij}[\phi^*;y]\frac{\partial}{\partial x_j}\int_\Lambda dz\,C_2^{leq,-1}(y,z)C_D(z,x)\right]\nonumber\\
&&=-\sum_i\frac{\partial}{\partial x_i}\left[\alpha_i[\phi^*;x]C_2^{leq}(x,y)\right]-\sum_i\frac{\partial}{\partial y_i}\left[\alpha_i[\phi^*;y]C_2^{leq}(x,y)\right]
\end{eqnarray}
where $\alpha$ is a $d$-dimensional vector 
\begin{equation}
\alpha[\phi;x]=\chi'[\phi;x]\chi^{-1}[\phi;x]J
\end{equation}
and we have considered that $\chi_{ij}[\phi;x]$ is a function that depends only on $\phi(x)$, that is $\chi_{ij}[\phi;x]=\chi_{ij}(\phi(x))$. Therefore $\chi'_{ij}[\phi;x]=\partial\chi_{ij}(u)/\partial u\vert_{u=\phi(x)}$.

The solution of this equation is very complex and it depends on the particular system and boundary conditions used. Let us work out explicitly a well known particular case: the pure {\it diffusive system} by taking:
\begin{equation}
V_0[\phi]=\int_\Lambda dx\,\left[v(\phi(x))-2E\cdot x \phi(x) \right]
\end{equation}
where $E$ is an external constant vector. With this choice we get:
\begin{equation}
G_i[\phi;x]=-\sum_j\left[D_{ij}[\phi;x]\partial_j\phi(x)-\chi_{ij}[\phi;x]E_j\right]
\end{equation}
where
\begin{equation}
D[\phi;x]=\frac{1}{2}v''(\phi(x))\chi[\phi;x]
\end{equation}
that is the so called {\it Einstein Relation}. We observe that in equilibrium (with the appropriate boundary conditions) we find that $\phi_{eq}^*(x)$ is solution of the {\it barometric equation}:
\begin{equation}
\nabla\phi_{eq}^*(x)=-\frac{2}{Vv'(\phi_{eq}^*(x))}E
\end{equation}
Moreover,
\begin{equation}
C_2^{eq}(x,y)=\frac{1}{v''(\phi_{eq}^*(x))}\delta(x-y)
\end{equation}
In a non equilibrium setup we obtain that the stationary state is solution of the equation:
\begin{equation}
-\sum_j\left[D_{ij}[\phi^*;x]\partial_j\phi^*(x)-\chi_{ij}[\phi^*;x]E_j\right]=J_i
\end{equation}
and the equation for $C_D$ is, in this case:
\begin{eqnarray}
&&\sum_{ij}\biggl[\frac{\partial}{\partial x_i}\left[\frac{\partial(D_{ij}[\phi^*;x]C_D(x,y))}{\partial x_j}-\chi_{ij}'[\phi^*;x]C_D(x,y)\right]\nonumber\\
&&+\frac{\partial}{\partial y_i}\left[\frac{\partial(D_{ij}[\phi^*;y]C_D(x,y))}{\partial y_j}-\chi_{ij}'[\phi^*;y]C_D(x,y)\right]\biggr]\nonumber\\
&&=\frac{1}{2}\left(\nabla\cdot\bar\alpha[\phi^*;x]\right)\delta(x-y)
\end{eqnarray}
where
\begin{equation}
\bar\alpha[\phi;x]=\chi'[\phi;x]D^{-1}[\phi;x]J
\end{equation}
In particular let us restrict to one dimension, $D=\text{cte}$, $E=0$ and $\chi[\phi;x]$ a positive second order polynomial of the form $\chi[\phi;x]=c_0+c_1\phi(x)+c_2\phi(x)^2$. We find  that
$J=-Dd\phi^*(x)/dx$.  That implies a stationary state: $\phi^*(x)=\phi^*(0)-Jx/D$, $J=D(\phi^*(L)-\phi^*(0))/L$, where we have fixed the values of $\phi$ at the boundaries of the segment $[0,L]$. Then, the correlation function is
\begin{equation}
C_2(x,y)=\frac{\chi[\phi^*;x]}{2D}\delta(x-y)+C_D(x,y)
\end{equation}
where the equation for $C_D$ is reduced to:
\begin{equation}
\left[\frac{d^2}{dx^2}+\frac{d^2}{dy^2}\right]C_D(x,y)=-2\frac{J^2}{D^3}c_2\delta(x-y)
\end{equation}
We can get an explicit solution for it  (see for instance Ref.\cite{Derrida})
\begin{equation}
C_D(x,y)=-\frac{J^2}{D^3}c_2\Delta^{-1}(x,y)
\end{equation}
with 
\begin{equation}
\Delta^{-1}(x,y)=-\frac{1}{L}\left[(L-x)y\theta(x-y)+x(L-y)\theta(y-x) \right]
\end{equation}
where $\theta(x)$ is the Heaviside function.
Observe the long range behavior of $C_D$ and that its sign  depends on the sign of $c_2$.

From this result  we can get the fluctuations of the averaged field $\rho[\phi]=\frac{1}{L}\int_{0}^L\, dx \phi(x)$ :

\begin{equation}
\Sigma\equiv\Omega\langle(\rho[\phi]-\rho^*)^2\rangle_{st}=\frac{1}{L^2}\int_{0}^L\,dx\int_{0}^L\,dy\, C_2(x,y)\equiv \Sigma_{leq}+\Sigma_D
\end{equation}
where
\begin{eqnarray}
\Sigma_{leq}&=&\frac{1}{2DL}\left[c_0+c_1\rho^*+\frac{c_2}{3}(\phi^*(0)^2+\phi^*(0)\phi^*(1)+\phi^*(1)^2)\right]\nonumber\\
\Sigma_D&=&\frac{c_2}{12DL}\left(\phi^*(0)-\phi^*(L)\right)^2
\end{eqnarray}
with  $\rho^*=\rho[\phi^*]$.
We see that the deviation from the local equilibrium is proportional to the square of the external gradient. This result has been derived  in the boundary driven Symmetric Simple Exclusion Process (SSEP) and in the Kipnis-Marchioro-Presutti model (KMP) \cite{Derrida,KMP,Bertini1}.

\item {\bf Bulk nonequilibrium:} Let us focus in a very simple nonequilibrium model at the bulk level that develops highly nontrivial correlations. Let 
\begin{equation}
G[\phi;x]=-D\nabla\phi
\end{equation}
where we assume that $D$ and $\chi$ are constant arbitrary d-dimensional matrices. One can easily  check that this system is time reversible if $D$ is proportional to $\chi$. Let us assume periodic boundary conditions such that $\phi^*(x)=cte$ is the stationary state. Therefore the currents on the system are zero: $G[\phi^*;x]=0$. 
The equation for $C_2$ is in this case given by:
\begin{equation}
\sum_{ij}D_{ij}\frac{\partial^2 \bar C_2(x-y)}{\partial x_i\partial x_j}=\frac{1}{2}\sum_{ij}\chi_{ij}\frac{\partial^2 \delta(x-y)}{\partial x_i\partial x_j}\label{gareq}
\end{equation} 
where we have assumed that the correlations are translational invariant: $C_2(x,y)=\bar C_2(x-y)$.  The solution of eq.(\ref{gareq}) is given by:
\begin{equation}
\bar C_2(u)=\int dk e^{iku}\hat C_2(k)\quad,\quad \hat C_2(k)=\frac{k\cdot \chi k}{k\cdot Dk}
\end{equation}
We see that $\hat C_2$ is non-analytic at $k=0$ when $D$ is not proportional to $\chi$ (this also implies $D$ and/or $\chi$ to be anisotropic) and $\bar C_2(u)$ has a  power law decay behavior \cite{Gar}. Let us remark the fact that this very simple  conservative dynamics with no macroscopic current has long range correlations just by breaking the proportionality between $D$ and $\chi$. We can think that the equilibrium is reached by a fine tuning of the system's parameters and that the normal behavior is the non-equilibrium one.

\item {\bf Strictly conserved models:} DD models  permits the average density
\begin{equation}
\frac{1}{\vert\Lambda\vert}\int_\Lambda dx\,\phi(x,t)=\bar\phi\quad\forall t
\end{equation}
 to be strictly conserved under suitable boundary conditions (for instance periodic). In this case, the stationary probability distribution can be written 
 \begin{equation}
 P_{st}[\phi]\simeq \exp\left[-\Omega V_0[\phi]\right]\delta\left(\int_\Lambda dx\,\phi(x)-\vert\Lambda\vert\bar\phi\right)\quad\Omega\rightarrow\infty\label{sc}
 \end{equation}
 where $V_0[\phi]$ is solution of the Hamilton-Jacobi equation:
 \begin{equation}
 H\left[\phi,\frac{\delta V_0[\phi]}{\delta\phi}\right]=0\label{HJ7}
 \end{equation}
 with $H$ given by eq.(\ref{hamDD}).
 That changes a bite the initial assumption we used from eqs.(\ref{star0},\ref{star1}) to derive the expressions for the correlations and we need to reformulate them. The Generating Functional (\ref{star2}) can be written 
 \begin{equation}
 Z[b]\propto \int D\phi\int ds\,\exp\left[-\Omega\mathcal{F}[\phi,b,s] \right]\quad\Omega\rightarrow\infty
 \end{equation}
 where
 \begin{equation}
 \mathcal{F}[\phi,b,s]=V_0[\phi]-\int_\Lambda dx b(x)\phi(x)+s\left(\int_\Lambda dx\, \phi(x)-\vert\Lambda\vert\bar\phi\right)
 \end{equation}
 The two body correlations are given by
 \begin{equation}
 C_2(x,y)=\lim_{\Omega\rightarrow\infty}\frac{\delta^2 I[b]}{\delta b(x)\delta b(y)}\biggr\vert_{b=0}
 \end{equation}
 with
 \begin{equation}
I[b]= \frac{1}{\Omega}\log Z[b]=-\mathcal{F}[\phi^*[b],b,s^*[b]]\label{II}
 \end{equation}
 and $(\phi^*[b],s^*[b])$ are the values that minimize $\mathcal{F}$:
 \begin{eqnarray}
 \frac{\delta V_0[\phi]}{\delta\phi(x)}\biggr\vert_{\phi=\phi^*[b]}&=&b(x)-s^*[b]\equiv\bar b(x)\nonumber\\
\int_\Lambda dx\,\phi^*(x;b)&=&\vert\Lambda\vert\bar\phi \quad\quad \forall\, b's\label{per}
 \end{eqnarray}
 We observe that $s^*[0]=0$ and we can get $\phi^*[b]$ and $s^*[b]$ from eqs.(\ref{per}) using a perturbative scheme around $\bar b=0$: 
 \begin{eqnarray}
 \phi^*(x;b)&=&\phi^*(x)+\int_\Lambda dy \,C_2^{NC}(x,y)\bar b(y)+O(\bar b^2)\nonumber\\
 s^*[b]&=&\frac{\displaystyle\int_\Lambda dx\int_\Lambda dy \,C_2^{NC}(x,y)\bar b(y) }{\displaystyle \int_\Lambda dx\int_\Lambda dy\, C_2^{NC}(x,y)}+O(\bar b^2)
 \end{eqnarray}
 where 
 \begin{equation}
 C_2^{NC}(x,y)=V_2^{-1}(x,y)
 \end{equation}
 and $V_2$ is given by eq.(\ref{V2}). That is, $C_2^{NC}$ is the two-body correlation function corresponding to a system with the same dynamics and boundaries as the original system but without the density conservation condition (observe that $V_0$ is computed using the Hamilton-Jacobi equation (\ref{HJ7}) where $\bar\phi$ is fixed when the stationary state is chosen).
 Finally we get $C_2$ by doing the expansion of $I[b]$, (\ref{II}), up to second order in $\bar b$ of their arguments:
 \begin{equation}
 C_2(x,y)= C_2^{NC}(x,y)-\frac{\displaystyle\int_\Lambda d\bar y \,C_2^{NC}(x,\bar y)\int_\Lambda d\bar x \,C_2^{NC}(\bar x,y) }{\displaystyle\int_\Lambda d\bar x \int_\Lambda d\bar y\,C_2^{NC}(\bar x,\bar y)}
 \end{equation} 
 We see that $\int_\Lambda dx\,C_2(x,y)=\int_\Lambda dy\,C_2(x,y)=0$ as expected. Observe also that the difference between $C_2$ and $C_2^{NC}$ is of order $\vert\Lambda\vert^{-1}$ and therefore both correlations coincide in the thermodynamic limit $\vert\Lambda\vert\rightarrow\infty$. At the practical level, the condition on the conservation of the density does not change the fact that first we should get $C_2^{NC}$ from the equation (\ref{DDcor}) as in the other cases and the system's physics  is completely characterized by it.
\end{itemize}

\subsection{An initial approach to define nonequilibrium dynamical ensembles}

We know from the equilibrium ensemble theory that there are several probability densities defined in the configurational space that give rise to the same macroscopic description in the thermodynamic limit \cite{Touchette}. For instance we are aware of the microcanonical, canonical and grand canonical ensembles. All the ensembles are characterized by the same Hamiltonian at the microscopic level and their differences are in the constraints they have: conservation of energy, of particles,....
This equivalence can be also transported to dynamic equations. For example we can build several stochastic dynamics such that all of them drive the system to the same equilibrium state.  At the mesoscopic level we have already seen that RD and DD systems at equilibrium have the same stationary distribution defined by the quasipotential $V_0[\phi]$ (except for a DD with strict conservation where   the conservation of the field appears explicitly and it affects, in a controlled manner, to the system correlations). In any case, all of them describe the same macroscopic state at the thermodynamic limit: their deterministic values and their correlations are equal.
 
 The equivalence of ensembles in nonequilibrium systems have been studied  from different point of views. Let us comment for example the comparison of turbulent Navier-Stokes hydrodynamic equations with different forcing and dissipative mechanisms  having them equal averages for any reasonable observable  \cite{Gallavotti0} or  
  the construction of equivalent {\it biased} dynamics at  the level of path's probabilities for Markov processes when it is included a constraint on an time averaged observable that give rise to the same large deviation distribution  \cite{Touchette0}.
  
  Here we question ourselves about the possibility to obtain a couple of RD and DD dynamics that drive a system to the same nonequilibrium stationary state.    
We assume that there are two levels of equivalence: (i) the strong equivalence where the quasipotential $V_0$ is the same in RD and DD models except for a conservation constraint  if necessary (see eq.(\ref{sc})) and (ii) the weak equivalence where we look for RD and DD models that have the same macroscopic stationary state and two body correlations.   
 To show the strong equivalence for a couple of dynamics is a nontrivial task because of the nonlocal structure of  $V_0$ and therefore its dependence on the boundary conditions and on the details of the dynamics. For this reason, it is more convenient to 
explore first the weak equivalence by  looking for the conditions  under which a RD and DD systems have equal two body correlations (observe that if we consider a DD with strict conservation of density we want to compare the RD correlation function with the DD's correlation $C_2^{NC}$). 
 
 The two body correlation is solution of eq.(\ref{co}) each one with their own hamiltonian. This implies eqs. (\ref{twobody}) and (\ref{DDcor}) for the RD and DD systems respectively. We do not know how to solve such equations in general except in the RD case when the $B$ kernel (\ref{dB}) is symmetric. That restricts a lot the types of nonequilibrium models we can study. Nevertheless it give us some interesting insight about the conditions for the existence of a weak equivalence.
 
 Let us assume a RD dynamics defined by a $F[\phi;x]$ and $h[\phi;x]$ functionals with the property:
\begin{equation}
B(x,y)\equiv\frac{h[\phi_1^*;y]}{h[\phi_1^*;x]}\frac{\delta F[\phi;x]}{\delta\phi(y)}\biggr\vert_{\phi=\phi_1^*}=B(y,x)\label{bker}
\end{equation} 
where $\phi_1^*$ is solution of $F[\phi_1^*;x]=0$.  We showed above that in this case the two body correlations are given by:
\begin{equation}
C_2(x,y)=-\frac{1}{2}h[\phi_1^*;x]h[\phi_1^*;y]B^{-1}(x,y)
\end{equation}
We impose that this correlation should be also solution of the eq. (\ref{DDcor}) for the two body correlations in the DD case (defined by $G[\phi;x]$ and $\chi_{ij}[\phi;x]$ functionals). Obviously the boundary conditions for $C_2$  are equal in both cases. In this way we get the relation:
\begin{equation}
h[\phi_1^*;y]\int_\Lambda dz\, K(x,z)h[\phi_1^*;z]B^{-1}(z,y)=\sum_{ij}\frac{\partial}{\partial x_i}\frac{\partial}{\partial y_j}\left[\chi_{ij}[\phi_2^*;x]\delta(x-y)\right]
\end{equation}
where 
\begin{equation}
K(x,y)=\frac{\delta}{\delta\phi(y)}\left(-\nabla\cdot G[\phi;x]\right)\biggr\vert_{\phi=\phi_2^*}
\end{equation}
with $\phi_2^*$ solution of $\nabla\cdot G[\phi_2^*;x]=0$.  After some trivial algebra we find that  the RD with $(F,h)$ and DD with $(G,\chi)$ have the same $C_2$ correlation provided that the following relationship is fulfilled:
\begin{equation}
\frac{\delta G_i[\phi;x]}{\delta\phi(y)}\biggr\vert_{\phi=\phi_2^*}=\sum_j\chi_{ij}[\phi_2^*;x]\frac{\partial}{\partial x_j}\left[\frac{1}{h[\phi_1^*;x]^2}\frac{\delta F[\phi;x]}{\delta\phi(y)}\biggr\vert_{\phi=\phi_1^*}\right]\label{pro}
\end{equation}

One can also check that 
\begin{equation}
G_i[\phi;x]=\sum_j\chi_{ij}[\phi;x]\frac{\partial}{\partial x_j}\left[\frac{F[\phi;x]}{h[\phi;x]^2}\right]\label{pro2}
\end{equation}
fulfills the relation (\ref{pro}). Observe that in this case the deterministic solutions for the RD and DD models coincide: $\phi_1^*=\phi_2^*=\phi^*$. In conclusion, we have shown the following property:

\begin{itemize}
\item {\it All RD models with a symmetric $B$-kernel (\ref{bker}) have an associated DD model given by (\ref{pro2}) such that both have the same macroscopic state and two body correlations and therefore they are, at least, weak-equivalent.}
\end{itemize}

We can apply this property to the equilibrium case where $F$ is of the form:
\begin{equation}
F[\phi;x]=-\frac{1}{2}h[\phi;x]^2 \frac{\delta V[\phi]}{\delta\phi(x)}
\end{equation}
for any arbitrary $h$ and $V$ functionals.  The B-kernel (\ref{pro2}) is in this case always symmetric and the weak equivalent conservative dynamics is the expected:
\begin{equation}
G_i[\phi;x]=-\frac{1}{2}\sum_j\chi_{ij}[\phi;x]\frac{\partial}{\partial x_j}\left[\frac{\delta V[\phi]}{\delta\phi(x)}\right]
\end{equation}
(see eq.(\ref{eq00})).
We know that  both dynamics  have the same quasipotential: $V_0[\phi]=V[\phi]$. That is,  they are also strong-equivalent. 

Let us see what happens for a simple RD system with nonequilibrium stationary state.  Assume that RD is defined by $(F,h)$ where
\begin{equation}
F[\phi;x]=-\frac{1}{2}g[\phi;x]^2 \frac{\delta V[\phi]}{\delta\phi(x)}
\end{equation}
with $V[\phi]$, $g[\phi;x]$ and $h[\phi;x]$ are given functionals. We know that whenever $g[\phi;x]\neq h[\phi;x]$ the system is in a nonequilibrium stationary state and the quasipotential $V_0[\phi]\neq V[\phi]$. The $B$-kernel (\ref{bker}) is in this case
\begin{equation}
B(x,y)=-\frac{1}{2}\frac{h[\phi^*;y]}{h[\phi^*;x]}g[\phi^*;x]^2\frac{\delta^2 V[\phi]}{\delta\phi(x)\delta\phi(y)}\biggr\vert_{\phi=\phi^*}
\end{equation} 
and it is symmetric when the deterministic solution is spatially homogeneous $\phi^*(x)=\phi^*=cte$. In such case we apply the result (\ref{pro2}) and the DD dynamics with the same two body correlations as the RD is:
\begin{equation}
G_i[\phi;x]=-\frac{1}{2}\sum_j\chi_{ij}[\phi;x]\frac{\partial}{\partial x_j}\left[\frac{g[\phi;x]^2}{h[\phi;x]^2}\frac{\delta V[\phi]}{\delta\phi(x)}\right]
\end{equation}
for any given $\chi_{ij}$. It is an open problem to see if there is a strong equivalence in this case.  We have just shown with this example that there exists weak equivalence between a family of RD and DD systems. It seems interesting to explore the possibility to find such equivalence in more complex dynamics. 

\section{Conclusions}

We have made an attempt to show general properties of nonequilibrium systems at stationary states assuming that they are described by continuum Langevin equations. We have studied the stationary measure at the small noise limit through the quasi-potential. This object is a natural extension of the Free Energy functional for systems at equilibrium. A quasipotential typically has a non-local structure, a strong dependence on the boundary conditions and also on the details of the dynamics and finally it may contain some non-differential behavior. These properties makes very diffucult to devise an apriori way of building them from some first principles as it occurs for the equilibrium. That prevents us to use the quasipotentials as a starting point to develop a hypothetical  nonequilibrium thermodynamics. 
We show how the most probable path to create a fluctuation from the stationary state is, in general and for nonequilibrium systems, different from the most probable path to relax from it. When both coincide the system is called {\it macroscopic time reversible}. This property is not enough to discriminate nonequilibrium stationary states from the equilibrium. In order to do that one should introduce  the {\it adjoint dynamics} as the Langevin equation for the time reversed process.  In this way it is defined equilibrium as the process in which the Langevin dynamics coincide with its adjoint.
The quasipotential near the stationary state is characterized by the stationary two-body correlations. We have explicitly built the equations to derive them, study several properties, examples and even perturbation schemes. That just to show the richness and nontrivial behavior of them.  

This work wants to show that there is a systematic way to  study  nonequilibrium systems from a theoretical point of view. It give us the possibility to study under the same scheme different non-equilibrium models and so to compare them to look for their common regularities.  Moreover, the advantage of this common theoretical ground is that any exact resolutions, approximation, perturbation scheme or assumption can be checked in many different scientific contexts from numerical models or real experiments. 

Finally, it is also observed that small changes in the overall functionals may imply large differences in the kind of results we derive from the theory. Therefore one of the main questions to be solved is to know the influence of the underlying microscopic details into the mesoscopic description. We know that in some important cases of boundary driven nonequilibrium systems (for example Fluctuating Hydrodynamics \cite{Fox}) the mesoscopic theory contains most of the necessary elements to describe correctly many observed phenomena. Nevertheless we would like to have in an ``a priori'' predictive way to connect safely the microscopic and mesoscopic descriptions.   

\section{Acknowledgements}
We thank to P.I. Hurtado and C.P. Espigares for very helpful comments. 
This work was supported by the Spanish governement project funded by MINECO/AEI/FEDERFIS2013-43201P and in part by AFOSR [grant FA-9550-16-1-0037].

\section*{APPENDIX I: From Langevin to Fokker Planck equations through a family of discretization schemes}
\subsection{RD case:}
Let us assume that the Langevin equation (\ref{lan1}) is the continuous limit of its time discrete version:
\begin{equation}
\phi(x,s+1)=\phi(x,s)+\epsilon\left[F[\phi;x,s,v]+h[\phi;x,s,v]\xi(x,s) \right] \label{ledd}
\end{equation}
where we also assume:
\begin{eqnarray}
F[\phi;x,s,v]&=&F[v\phi(x,s)+(1-v)\phi(x,s+1);x]\nonumber\\
h[\phi;x,s,v]&=&h(v\phi(x,s)+(1-v)\phi(x,s+1))
\end{eqnarray}
with $v\in[0,1]$, $x\in \Lambda \subset {\rm I\!R}^d$, $s\in Z$ and $F[\phi;x]$   is a  functional on $\phi(y)$'s with $y$ pertaining to a finite open region around $x$ while $h(\lambda)$ is a function. The random field $\xi$ is a Gaussian white noise characterized by:
\begin{equation}
\langle\xi(x,s)\rangle=0\quad ,\quad \langle\xi(x,s)\xi(x',s')\rangle=\frac{1}{\epsilon\Omega}\delta(x,x')\delta_{s,s'}
\end{equation}
Observe that for any value $v\in[0,1]$ the limit $\epsilon\rightarrow 0$ of this discrete equation gives rise to the continuous Langevin equation (\ref{lan1}). 

We can expand the Langevin equation (\ref{ledd}) in powers of $\epsilon$:
\begin{eqnarray}
\phi(x,s+1)&=&\phi(x,s)+\epsilon h(\phi(x,s))\xi(x,s)+\epsilon F[\phi;x,s]\nonumber\\
&+&(1-v)\epsilon^2h(\phi(x,s))h'(\phi(x,s))\xi(x,s)^2+O(\epsilon^{3/2})
\end{eqnarray}
where we have assumed that $\xi$ is of order $\epsilon^{-1/2}$.

The probability to find a given configuration $\phi$ at time $s$ is defined by
\begin{equation}
P[\phi;s+1]=\langle\prod_{x\in\Lambda}\delta(\phi(x)-\phi(x,s+1))\rangle_{\xi}\label{PP}
\end{equation}
where $\phi(x,s)$ is the solution of the Langevin equation for a given random noise realization and $\langle\cdot\rangle_\xi$ is the average over all noise realizations with their corresponding Gaussian weight. We can substitute the $\epsilon$ expanded Langevin equation into eq.(\ref{PP}) and after some algebraic manipulation we get
\begin{eqnarray}
P[\phi;s+1]&=&\int \prod_{x\in\Lambda}\left[d\bar\phi(x)\right] P[\bar\phi;s]
\langle\prod_{x\in\Lambda}\delta\biggl(\phi(x)-\bar\phi(x)-\epsilon h(\bar\phi(x))\xi(x,s)-\epsilon F[\bar\phi;x]\nonumber\\
&-&(1-v)\epsilon^2h(\bar\phi(x))h'(\bar\phi(x))\xi(x,s)^2+O(\epsilon^{3/2})\biggr)\rangle_\xi \label{FPD0}
\end{eqnarray}
We have used the fact that $\phi(x,s)$ only depend on $\xi$'s of previous times, $s'<s$, and therefore we can break the averages over $\xi$'s.
We expand the last expression with respect $\epsilon$ by using the perturbative formula 
\begin{eqnarray}
\prod_{n}\delta(a(n)&+&b(n)\eta+c(n)\eta^2)=\left(\prod_n\delta(a(n))\right)+\eta\sum_m\left(\prod_{n\neq m}\delta(a(n))\right)\delta'(a(m))b(m)\nonumber\\
&&+\frac{1}{2}\eta^2\sum_m\biggl[\left(\prod_{n\neq m}\delta(a(n))\right)\biggl[\delta''(a(m))b(m)^2+2\delta'(a(m))c(m)\biggr]\nonumber\\
&&+\sum_{m'\neq m}\left(\prod_{n\neq m,m'}\delta(a(n))\right)\delta'(a(m))\delta'(a(m'))b(m)b(m')
\biggr]+O(\eta^3)\label{for}
\end{eqnarray}
that we get by doing the first two derivatives with respect $\eta$ on the right hand side of eq.(\ref{for}) and  expanding the remaining expression up to second order in $\eta$. In our case we identify $\eta=\epsilon^{1/2}$. 

Finally, we can do the averages over $\xi$'s and we get (in the limit $\epsilon\rightarrow 0$) the Fokker-Planck equation:
\begin{eqnarray}
\partial_t P[\phi;t]&=&\int_{\Lambda}dx\frac{\delta}{\delta\phi(x)}\biggl[-(F[\phi;x]
+\frac{(1-v)}{\Omega}h(\phi(x))h'(\phi(x)))P[\phi;t]\nonumber\\
&+&\frac{1}{2\Omega}\frac{\delta}{\delta\phi(x)}\left(h(\phi(x))^2P[\phi;t]\right)
\biggr]
\end{eqnarray}
 One can show  that the observables (averages) computed with this Lagrangian do not depend on the $v$ used \cite{Les}.
For $v=1$ (Ito's discretization) we obtain the Fokker-Planck equation that it is used in this paper:
\begin{equation}
\partial_t P[\phi;t]=\int_\Lambda dx\frac{\delta}{\delta\phi(x,t)}\left[-F[\phi;x,t]P[\phi;t]+\frac{1}{2\Omega}\frac{\delta}{\delta\phi(x,t)}\left(h[\phi;x,t]^2P[\phi;t]\right) \right]\label{fp0}
\end{equation} 

\subsection{DD case:}
In this case it is necessary to define an space and time discretizations. The field at lattice site $n\in Z^d$ at discrete time $s\in Z$, $\phi(n,s)$, is solution of the discrete Langevin equation:
\begin{equation}
\phi(n,s+1)=\phi(n,s)-\frac{\epsilon}{2a}\sum_{\alpha=1}^d\left[j_\alpha(\phi;n+i_\alpha,s)-j_\alpha(\phi;n-i_\alpha,s)\right]
\end{equation}
where $i_\alpha$ is the unit vector in the direction $\alpha$ and
\begin{equation}
j_\alpha(\phi;n,s)=G_\alpha[\phi;n,s]+\sum_{\beta=1}^d\sigma_{\alpha\beta}[\phi;n,s]\psi_\beta(n,s)
\end{equation}
\begin{equation}
\langle\psi_\alpha(n,s)\psi_\beta(n',s')\rangle=\frac{1}{\tilde\Omega\epsilon a^d}\delta_{\alpha,\beta}\delta_{n,n'}\delta_{s,s'}
\end{equation}
where $a$ and $\epsilon$ are the lattice node separation in space and time respectively. For simplicity we are considering just the Ito's scheme.

The probability to find a given configuration $\phi$ at time $s$ is defined by
\begin{equation}
P[\phi;s+1]=\langle\prod_{n\in\Lambda}\delta(\phi(n)-\phi(n,s+1))\rangle_{\psi}
\end{equation}
where $\phi(n,s)$ is the solution of the Langevin equation for a given random noise realization, that is, it depends on $\psi$ and $\langle\cdot\rangle_\psi$ is the average over all noise realizations with their corresponding Gaussian weight.
We can insert the right hand side of the Langevin equation and we introduce an auxiliary field $\bar\phi$:
\begin{eqnarray}
P[\phi;s+1]&=&\langle\int \prod_{n\in\Lambda}\left[d\bar\phi(n)\delta(\bar\phi(n)-\phi(n,s)) \right]\nonumber\\
&&\prod_{n\in\Lambda}\delta\left(\phi(n)-\bar\phi(n)+\frac{\epsilon}{2a}\sum_{\alpha=1}^d\left[j_\alpha(\bar\phi;n+i_\alpha)-j_\alpha(\bar\phi;n-i_\alpha)\right]\right)\rangle_\psi
\end{eqnarray}
We use the fact that  the noise $\psi$ is time uncorrelated and  the Ito's prescription. Moreover, $\phi(n,s)$ only depend on $\psi$'s with times strictly smaller than $s$. Therefore we can break the average over $\psi$ and we get:
\begin{eqnarray}
P[\phi;s+1]&=&\int \prod_{n\in\Lambda}\left[d\bar\phi(n)\right] P[\bar\phi;s]\nonumber\\
&&\langle\prod_{n\in\Lambda}\delta\left(\phi(n)-\bar\phi(n)+\frac{\epsilon}{2a}\sum_{\alpha=1}^d\left[j_\alpha(\bar\phi;n+i_\alpha)-j_\alpha(\bar\phi;n-i_\alpha)\right]\right)\rangle_\psi\label{FPD}
\end{eqnarray}
We expand the last expression for $\epsilon\leq 1$ taking into account that $\psi$ is of order $\epsilon^{-1/2}$. We can use the formula (\ref{for}) with
\begin{eqnarray}
a(n)&=&\phi(n)-\bar\phi(n)\nonumber\\
b(n)&=&\frac{\epsilon^{1/2}}{2 a}\sum_{\alpha=1}^d\sum_{\beta=1}^d\left[\sigma_{\alpha\beta}[\bar\phi;n+i_{\alpha}]\psi_{\beta}(n+i_\alpha,s)-\sigma_{\alpha\beta}[\bar\phi;n-i_{\alpha}]\psi_{\beta}(n-i_\alpha,s) \right]\nonumber\\
c(n)&=&\frac{1}{2a}\sum_{\alpha=1}^d\left[G_\alpha[\bar\phi;n+i_\alpha]-G_\alpha[\bar\phi;n-i_\alpha] \right]
\end{eqnarray}
After substituting this expansion into eq.(\ref{FPD}) we can do explicitly the averages over $\psi$ and after some algebra we get
\begin{eqnarray}
P[\phi;s+1]&=&P[\phi;s]+\epsilon\sum_{m\in\Lambda}\frac{\partial}{\partial\phi(m)}\biggl[P[\phi;s]\frac{1}{2a}\sum_{\alpha=1}^d\left(G_\alpha[\phi;n+i_\alpha]-G_\alpha[\phi;n-i_\alpha] \right)\nonumber\\
&+&\frac{1}{8\tilde\Omega a^{d+2}}\sum_{\alpha=1}^d\sum_{\beta=1}^d\biggl(\frac{\partial}{\partial\phi(m+i_\alpha-i_\beta)}\left(P[\phi;s]\chi_{\alpha\beta}[\phi;m+i_\alpha]\right)\nonumber\\
&-&\frac{\partial}{\partial\phi(m+i_\alpha+i_\beta)}\left(P[\phi;s]\chi_{\alpha\beta}[\phi;m+i_\alpha]\right)\nonumber\\
&-&\frac{\partial}{\partial\phi(m-i_\alpha-i_\beta)}\left(P[\phi;s]\chi_{\alpha\beta}[\phi;m-i_\alpha]\right)\nonumber\\
&+&\frac{\partial}{\partial\phi(m-i_\alpha+i_\beta)}\left(P[\phi;s]\chi_{\alpha\beta}[\phi;m-i_\alpha]\right)
\biggr)
\biggr]+O(\epsilon^2)
\end{eqnarray}
where
\begin{equation}
\chi_{\alpha\beta}[\phi;n]=\sum_{\gamma=1}^d\sigma_{\alpha\gamma}[\phi;n]\sigma_{\beta\gamma}[\phi;n]
\end{equation}
This expression can be written in a more compact form by using the definition:
\begin{equation}
\left(\partial_\alpha\frac{\partial}{\partial\phi(n)}\right)\equiv\frac{1}{2a}\left(\frac{\partial}{\partial\phi(n+i_\alpha)}-\frac{\partial}{\partial\phi(n-i_\alpha)}\right)
\end{equation}
Therefore we get
\begin{eqnarray}
\frac{1}{\epsilon}\left[P[\phi;s+1]-P[\phi;s]\right]&=&\sum_{\alpha=1}^d\sum_{m\in\Lambda}\left(\partial_\alpha\frac{\partial}{\partial\phi(m)}\right)\biggl[-G_\alpha[\phi;m]P[\phi;s]\nonumber\\
&+&\frac{1}{2\tilde\Omega a^d}\sum_{\beta=1}^d\left(\partial_\beta\frac{\partial}{\partial\phi(m)}\right)\left(\chi_{\alpha\beta}[\phi;m]P[\phi;s]\right)
\biggr]+O(\epsilon)
\end{eqnarray}
where we have used the properties:
\begin{eqnarray}
\sum_{m\in\Lambda}\frac{\partial}{\partial\phi(m)}\left(\partial_\alpha G[\phi;m]\right)&=&-\sum_{m\in\Lambda}\left(\partial_\alpha\frac{\partial}{\partial\phi(m)}\right)G[\phi;m]\nonumber\\
\left(\partial_\alpha\frac{\partial}{\partial\phi(m)}\right)(Q[\phi]F(\phi(m)))&=&F(\phi(m))\partial_\alpha\left(\frac{\partial Q[\phi]}{\partial\phi(m)}\right)
\end{eqnarray}
with $F(\lambda)$ being a function.

In the limit $\epsilon\rightarrow 0$ and $a\rightarrow 0$ and defining $\tilde\Omega=a^d\Omega$  we recover the Fokker-Planck equation for diffusive systems:
\begin{eqnarray}
\partial_t P[\phi;t]&=&\int_\Lambda dx\sum_{\alpha=1}^d\left(\partial_\alpha\frac{\delta}{\delta\phi(x,t)}\right)\biggl[ -G_{\alpha}[\phi;x,t]P[\phi;t]\nonumber\\
&+&\frac{1}{2\Omega}\sum_{\beta=1}^d\left(\partial_\beta\frac{\delta}{\delta\phi(x,t)}\right)\left(\chi_{\alpha,\beta}[\phi;x,t]P[\phi;t]\right)\biggr]\label{fp1}
\end{eqnarray}
where
\begin{equation}
\chi_{\alpha,\beta}[\phi;x,t]=\sum_{\gamma=1}^d\sigma_{\alpha,\gamma}[\phi;x,t]\sigma_{\beta,\gamma}[\phi;x,t]
\end{equation}
and 
\begin{equation}
\left(\partial_\alpha\frac{\delta}{\delta\phi(x)}\right)=\lim_{a\rightarrow 0}\frac{1}{2a}\left(\frac{\partial}{\partial\phi(n+i_\alpha)}-\frac{\partial}{\partial\phi(n-i_\alpha)}\right)
\end{equation}
where $x=na$. This operator has the useful property
\begin{equation}
\left(\partial_\alpha\frac{\delta}{\delta\phi(x,t)}\right)H[\phi;x,t]=\partial_\alpha\left(\frac{\delta}{\delta\phi(x,t)}H[\phi;x,t]\right)-\frac{\delta}{\delta\phi(x,t)}\left(\partial_\alpha H[\phi;x,t]\right)
\end{equation}

\section*{APPENDIX II: Large Deviations and Green-Kubo relations}

Let us define the bulk average of an observable $a[\phi;x,t]$ for a given time $t$ :
\begin{equation}
a[\phi;t]=\frac{1}{\Lambda}\int_\Lambda dx\, a[\phi;x,t]
\end{equation}
Its time average over the time interval $[0,T]$ is then
\begin{equation}
a_T[\phi]=\frac{1}{T}\int_{0}^T dt\, a[\phi;t]
\end{equation}
If the stochastic model is well behaved we can apply the Law of Large Numbers  in the sense that
\begin{equation}
a^*\equiv\langle a[\phi;0]\rangle_{ss}= \lim_{T\rightarrow\infty}a_T[\phi]
\end{equation}
The probability to observe a certain value of $a_T[\phi]=a$  assuming that at time $t=0$ the system is at the stationary state is given by:
\begin{equation}
P(a;T)=\int D\phi[0,T]P_{ss}[\phi(0)]P[\{\phi\}[0,T]]\delta(a-a_T[\phi])\label{LD7}
\end{equation}
The Large Deviation Principle states that for large values of $T$ this distribution should be very peaked around $a^*$. In fact  in such limit it should be of the form:
\begin{equation}
P(a;T)\simeq e^{-T R(a)}\quad\quad T\rightarrow\infty
\end{equation}
with
\begin{equation}
R(a^*)=0\quad\quad R'(a^*)=0
\end{equation}
Therefore
\begin{equation}
\lim_{T\rightarrow\infty}T \left\langle (a-a^*)^2\right\rangle_P=R''(a^*)^{-1}
\end{equation}
where $\langle\cdot\rangle_P$ means the the average is done with the $P(a;T)$ distribution. We can now substitute $P(a,T)$ by its path definition and we get the Green-Kubo relation:
\begin{equation}
\frac{1}{2R''(a^*)}=\int_{0}^\infty d\tau\,\langle (a[\phi;0]-a^*)(a[\phi;\tau]-a^*)\rangle
\end{equation}
where now $\langle\cdot\rangle$ is the path average defined above. 

We can apply this scheme to our RD and DD models and obtain (for a given $a[\phi;x,t]$) the function $R(a)$. As an example, let us just study a  RD system with 
\begin{eqnarray}
a[\phi;x,t]&\rightarrow&\phi(x,t)\nonumber\\ 
a[\phi;t]&\rightarrow&\rho[\phi;t]=\frac{1}{\Lambda}\int_\Lambda dx\, \phi(x,t) \nonumber\\
a_T[\phi]&\rightarrow&\rho_T[\phi]=\frac{1}{T}\int_0^T dt\,\rho[\phi;t]
\end{eqnarray}
and $P[\{\phi\}[0,T]]$ is given by eq.(\ref{path}). The probability to observe a given average density over the space and a time interval $T$ at the stationary state, $\rho_T[\phi]=\rho$, is:
\begin{equation}
P[\rho;T]=\int D\phi[0,T]P_{ss}[\phi(0)]\int_{c-i\infty}^{c+i\infty}\frac{d\lambda}{2\pi i}\exp\left[-\Omega T R[\{\phi\}[0,T]]\right]
\end{equation}
where
\begin{equation}
R[\{\phi\}[0,T],\lambda]=\frac{1}{2T}\int_{0}^{T}dt\int_{\Lambda}dx\left(\frac{\partial_t\phi(x,t)-F[\phi;x,t]}{h[\phi;x,t]}\right)^2+\lambda\left(\rho-\rho_T[\phi]\right)
\end{equation}
and we have used in eq. (\ref{LD7}) the representation of the Dirac delta by the integral on $\lambda$ . We can compute explicitly $P[\rho;T]$ when $T\rightarrow\infty$  because the integrals are dominated by its minimum value over the fields and $\lambda$. That is
\begin{equation}
P[\rho,T]\simeq \exp[-\Omega TR[\{\tilde\phi\}[0,T],\tilde\lambda]]
\end{equation}
where $\tilde\phi$ and $\tilde\lambda$ are solutions of
\begin{equation}
\frac{\delta R}{\delta\phi(y,\tau)}\biggr\vert_{\phi=\tilde\phi,\lambda=\tilde\lambda}=0\quad,\quad \frac{\partial R}{\partial\lambda}\biggr\vert_{\phi=\tilde\phi,\lambda=\tilde\lambda}=0
\end{equation}
In general, these set of equations have many different type of solutions (see for instance \cite{Hurtado}), static and dynamics that are local extremals of $R$. It is a daunting practical  task  to get some solutions and to check which is the one that is the absolute minimum for $R$. Let us assume the simplest case in which the deterministic solution of the Langevin equation is constant in space: $\phi^*(x,t)=\rho^*=cte$. Obviously, when $\rho=\rho^*$ we expect that $\tilde\phi(x,t)=\rho^*$. For values of $\rho$ near the stationary state solution $\rho^*$ we can assume by continuity that $\tilde\phi(x,t)=\rho$ is a solution of the equations. This  ansatz is the so-called {\it additivity principle} \cite{Bodineau}. 
In this case
\begin{equation}
R[\{\tilde\phi\}[0,T],\tilde\lambda]=\Lambda\frac{F[\rho]^2}{2h[\rho]^2}\equiv R[\rho]
\end{equation}
Therefore
\begin{equation}
R''[\rho^*]=\Lambda \frac{F'[\rho^*]^2}{h[\rho^*]}
\end{equation}
and the Green-Kubo relation is
\begin{equation}
\frac{h[\rho^*]^2}{F'[\rho^*]^2}=2\Omega\int_{R^d}dx\,\int_0^\infty d\tau\,\langle(\phi(0,0)-\phi^*)(\phi(x,\tau)-\phi^*)\rangle
\end{equation}
in the limit $\Lambda\rightarrow\infty$ and assuming spatial translation invariance.

We can study in the same way different observables. In the DD case it has been studied extensively the time averaged mean current in some 1-d systems \cite{Bertini2}:
\begin{equation}
J_T[\phi]=\frac{1}{T\Lambda}\int_0^T dt\,\int_\Lambda dx\, j(x,t)
\end{equation}
In one dimension it is shown that the additivity principle is correct when we look for fluctuations of the current near the stationary value but, in general, it fails for large current fluctuations where the solutions that minimize the functional $R$ are much more complex that the uniform solution. For instance, this occurs when we use periodic boundary conditions where such solutions are soliton-like functions that move around the system at constant speed. Moreover, it has been shown that in two dimensions the KMP model \cite{KMP} with a thermal gradient in one direction and periodic boundary conditions in the other, presents a solution ({\it weak additivity principle}) that is not spatially uniform but is a better minimizer than the uniform solution even near the stationary value \cite{weak}.  

\section*{APPENDIX III: The method of characteristics to solve Hamilton-Jacobi equations}
We just reproduce the page 233 in Gallavotti's book {\it Elements of Mechanics} \cite{Gall}. Let $S(q,t)$ to be solution of the Hamilton-Jacobi equation
\begin{equation}
H(\frac{\partial S(q,t)}{\partial q},q,t)+\frac{\partial S(q,t)}{\partial t}=0\label{e0}
\end{equation} 
where $H=H(p,q,t)$ is a given function on its arguments. Let us assume the following differential equation:
\begin{equation}
\frac{dq}{dt}=\frac{\partial H}{\partial p}\biggr\vert_{p=\frac{\partial S}{\partial q}} \label{e1}
\end{equation}
with the initial condition $q(t_0)=q_0$. Then, we can show that if we take
\begin{equation}
p(t)=\frac{\partial S}{\partial q}\biggr\vert_{q=q(t)}
\end{equation}
with $q(t)$ solution of eq.(\ref{e1}), then the functions $(q(t),p(t))$ are {\it solutions} of the Hamilton equations with  Hamiltonian $H(p,q,t)$ and initial values:
$q(t_0)=q_0$ and $p(t_0)=\partial S/\partial q\vert_{q=q_0}$. That is, each solution of the Hamilton-Jacobi equation (\ref{e0}) corresponds to a Hamiltonian dynamics.

In order to show this assertion we just check that $p(t)$ so defined is solution  of the corresponding Hamilton equation: $dp/dt=-\partial H/\partial q$:
\begin{equation}
\frac{dp_i}{dt}=\frac{d}{dt}\left(\frac{\partial S}{\partial q_i}\biggr\vert_{q=q(t)}\right)=\sum_j \frac{\partial^2 S}{\partial q_i\partial q_j}\frac{dq_j}{dt}+\frac{\partial^2 S}{\partial t\partial q_i}\label{e3}
\end{equation}
but deriving the Hamilton-Jacobi equation by $\partial/\partial q_i$ we find the relation:
\begin{equation}
\sum_j\frac{\partial^2 S}{\partial q_i\partial q_j}\frac{dq_j}{dt}+\frac{\partial H(p,q,t)}{\partial q_i}\biggr\vert_{p=\frac{\partial S}{\partial q}}+\frac{\partial^2 S}{\partial t\partial q_i}=0
\end{equation}
that we can use in eq.(\ref{e3}) to get the desired result:
\begin{equation}
\frac{dp_i}{dt}=-\frac{\partial H(p,q,t)}{\partial q_i}\quad , \quad \frac{dq_i}{dt}=\frac{\partial H(p,q,t)}{\partial p_i}
\end{equation}
with the above metioned initial conditions. In the case of a time independent hamiltonian $S(q,t)=W(q)-\alpha t$, where $\alpha$ is a constant fixed at initial time. 

We can find $S(q,t)$ just by studying the time behavior of $S(q(t),t)$ with $q(t)$ solution of the Hamilton equations:
\begin{equation}
\frac{dS(q(t),t)}{dt}=\sum_i \frac{\partial S(q,t)}{\partial q_i}\biggr\vert_{q=q(t)}\frac{dq_i}{dt}+\frac{\partial S(q,t)}{\partial t}\biggr\vert_{q=q(t)}
\end{equation}
We do a time integration to it and we get:
\begin{equation}
S(q(t),t)-S(q(t_0),t_0)=\int_{t_0}^t d\tau\sum_i p_i(\tau)\frac{dq_i(\tau)}{d\tau}+\int_{t_0}^td\tau\frac{\partial S(q,\tau)}{\partial \tau}\biggr\vert_{q=q(\tau)}
\end{equation} 
where $q((\tau),p(\tau))$ are the solutions of the Hamilton equations with initial conditions $q(t_0)=q_0$ and  $p(t_0)=\partial S/\partial q\vert_{q=q_0}$. It is convenient to choose: $p(t_0)=0$, that is, the value of $q_0=q^*$ in which $S(q,t_0)$ has an extreme: $\partial S/\partial q \vert_{q=q^*}=0$.

\section*{APPENDIX IV: Path integral method to obtain the correlations}

Let us study the RD case as an example. In order to obtain $V_0[\phi]$ from the path integral formalism we have to solve the evolution equations for $(\bar\phi(x,t),\pi(x,t))$ given by eqs. (\ref{HJ}) with boundary conditions $(\bar\phi(x,-\infty),\pi(x,-\infty))=(\phi^*(x),0)$ and $\bar\phi(x,0)=\phi(x)$. The quasi potential is obtained from eq.(\ref{action2}) by solving Hamilton's eqs. (\ref{Heq}) with the RD Hamiltonian given by  (\ref{hamRD}). We know that the two body correlation $C_2(x,y)$ is related with the second derivative of the quasi-potential at the deterministic stationary state (whenever $V_0$ is differentiable at such point). Therefore we want to solve Hamilton's equations when $\bar\phi(x,0)=\phi^*(x)+\Omega^{-1/2}\omega(x)$. We have no a priori guarantee that there are many paths that connect the initial condition $\phi^*$ at time $-\infty$ to a small deviation from it, $\bar\phi$. Moreover, It could also be that the path that minimizes the Lagrangian functional is one whose trajectory makes tours far from the initial point. An analytic solution for such types of situations are far from our actual capabilities.
 Let us focus then on the simple assumption that the linearized dynamics approximates correctly the path that connects the initial state with the final perturbed one. As we will see, this assumption is, in practice equivalent to the local differentiability  of the quasi-potential. 

Let us linearize the evolution eqs. (\ref{HJ}) assuming:
\begin{equation}
\bar\phi(x,t)=\phi^*(x)+\frac{1}{\sqrt{\Omega}}h[\phi^*;x]\bar\omega(x,t)\quad,\quad \pi(x,t)=\frac{1}{\sqrt{\Omega}\,h[\phi^*;x]}\bar\eta(x,t)
\end{equation}
then
\begin{eqnarray}
\partial_t \bar\omega(x,t)&=&\int_\Lambda dy B(x,y)\bar\omega(y,t)+\bar\eta(x,t)\nonumber\\
\partial_t \bar\eta(x,t)&=&-\int_{\Lambda} dy B(y,x)\bar\eta(y,t)
\end{eqnarray}
where $B(x,y)$ is defined in eq.(\ref{dB}). And the initial conditions are: $(\bar\omega(x,-\infty),\bar\eta(x,-\infty))=(0,0)$ and $\bar\omega(x,0)=\omega(x,0)$.
The quasi-potential is, in this approximation given by:
\begin{equation}
V_0[\phi]=V_0[\phi^*]+\frac{1}{2\Omega}\int_{-\infty}^0dt\,\int_\Lambda dx\,\bar\eta(x,t)^2
\end{equation}
let us remark that the trajectory $\bar\eta(x,t)$ contains the boundary conditions and therefore the $\omega(x)=\sqrt{\Omega}(\phi(x)-\phi^*(x))$ field.

\begin{eqnarray}
\partial_t \bar\omega&=&B\bar\omega+\bar\eta\nonumber\\
\partial_t\bar\eta&=&-B^T\eta
\end{eqnarray}
where $\bar\epsilon$ and $\bar\eta$ are vectors and $B$ a matrix and $B^T$ its transposed. The general solution is then
\begin{eqnarray}
\bar\eta(t)&=&e^{-tB^T}\bar\eta_0\nonumber\\
\bar\omega(t)&=&e^{tB}a_0+\int^td\tau\,e^{(t-\tau)B}e^{-\tau B^T}\bar\eta_0
\end{eqnarray}
where $\bar\eta_0$ and $a_0$ are constant vectors to be determined. First we assume that $\bar\omega(0)=\omega$, then
\begin{equation}
\omega=a_0+C(0)\bar\eta_0\quad,\quad C(t)=\int^td\tau\,e^{-\tau B}e^{-\tau B^T}
\end{equation}
where
\begin{equation}
\bar\eta_0=C(0)^{-1}(\omega-a_0)
\end{equation}
Now we assume that $B$ can be diagonalized  that is, there exists a $Q$ matrix such that $B=QDQ^{-1}$ with $D_{ij}=\lambda_i\delta_{i,j}$, $Q_{ij}=v_i(\lambda_j)$ where $(\lambda, v(\lambda))$ are the right eigenvalues and eigenvectors of $B$: $Bv(\lambda)=\lambda v(\lambda)$, $Q^{-1}_{ij}=w_j^*(\lambda_i)$ where $(\lambda^*, w(\lambda))$ are the left eigenvalues and eigenvectors of $B$:$B^Tw(\lambda)=\lambda^*w(\lambda)$ ($a^*$ stands for the complex conjugate of $a$). Notice that the set of eigenvalues of $B$ and $B^T$ are the same. Two useful orthogonal properties can be derived from $QQ^{-1}=Q^{-1}Q=1$: 
\begin{equation}
w^*(\lambda_i)\cdot v(\lambda_j)=\delta_{i,j}\quad,\quad \sum_k w_i^*(\lambda_k)v_j(\lambda_k)=\delta_{i,j}
\end{equation}
Observe that if $B$ is non-symmetric the set of eigenvectors may not be an orthonormal vector base.

With all these information we may introduce the boundary conditions to our general solutions. First we see that
\begin{equation}
\bar\eta(t)=(Q^{-1})^Te^{-tD}Q^T\bar\eta_0 
\end{equation}
we know that $\bar\eta(-\infty)=0$ implying that the real part of all the eigenvalues of $B$ should be negative:
\begin{equation}
Re(\lambda_i)<0\quad \forall i
\end{equation}
this is a ``stability condition'' over the dynamics and it is equivalent to ask that arbitrary and small perturbation to the deterministic stationary state will relax to it.
The second condition is $\bar\omega(-\infty)=0$. Let us write $\bar\omega(t)$ solution in function of its eigenvalues:
\begin{equation}
\bar\omega(t)=Qe^{tD}Q^{-1}a_0+\int^td\tau\,Qe^{(t-\tau)D}Q^{-1}(Q^{-1})^Te^{-\tau D}Q^T\bar\eta_0
\end{equation}
First we can show that the integral term tends to zero when $t\rightarrow-\infty$ because:
\begin{equation}
(\int^t d\tau e^{(t-\tau)D}Q^{-1}(Q^{-1})^Te^{-\tau D})_{ij}=-\frac{(Q^{-1}(Q^{-1})^T)_{ij}}{\lambda_i+\lambda_j}e^{-t\lambda_j}
\end{equation}
and we are assuming $Re(\lambda_i)<0$ $\forall i$. In the other hand the first term always diverge when applied to nonzero $a_0$ when $t\rightarrow\infty$. Therefore $a_0=0$ and the solution compatible with the boundary conditions is:
\begin{equation}
\bar\eta(t)=(Q^{-1})^Te^{-tD}Q^TC(0)^{-1}\omega\quad,\quad \bar\omega(t)=Qe^{Dt}Q^{-1}C(t)C(0)^{-1}\omega
\end{equation}
where
\begin{equation}
C(t)_{ij}=-\sum_{ks}Q_{ik}(Q^{-1}(Q^{-1})^T)_{ks}(Q^T)_{sj}\frac{e^{-(\lambda_k+\lambda_s)t}}{\lambda_k+\lambda_s}
\end{equation}
Finally, the quasi potential is:
\begin{equation}
\Omega V_0[\phi]=\frac{1}{2}\omega^T(C(0)^{-1})^T\omega
\end{equation}
and the two body correlation is
\begin{equation}
\bar C=C(0)^T
\end{equation}
that in the continuum limit reproduces equation (\ref{corr}).

\end{document}